\documentclass[aip,amsmath,amssymb,reprint,]{revtex4-1}

\usepackage{graphicx}
\usepackage{dcolumn}
\usepackage{bm}

\usepackage[utf8]{inputenc}
\usepackage[T1]{fontenc}
\usepackage{mathptmx}
\usepackage{etoolbox}
\usepackage{xcolor}
\usepackage{hyperref}

\usepackage{verbatim}

\makeatletter
\def\@email#1#2{%
 \endgroup
 \patchcmd{\titleblock@produce}
  {\frontmatter@RRAPformat}
  {\frontmatter@RRAPformat{\produce@RRAP{*#1\href{mailto:#2}{#2}}}\frontmatter@RRAPformat}
  {}{}
}%
\makeatother

\begin{document}


\preprint{AIP/123-QED}


\title[Low Energy Phonon Bursts Created By Fast Neutron Damage]{Low Energy Phonon Bursts Created By Fast Neutron Damage}

\author{A.~Armatol} \affiliation{Univ. Lyon, Universit\'e Lyon 1, CNRS/IN2P3, IP2I-Lyon, F-69622, Villeurbanne, France}
\author{C. Augier} \affiliation{Univ. Lyon, Universit\'e Lyon 1, CNRS/IN2P3, IP2I-Lyon, F-69622, Villeurbanne, France}
\author{L.~Berg\'e}\affiliation{Universit\'e Paris-Saclay, CNRS/IN2P3, IJCLab, 91405 Orsay, France}
\author{J.~Billard} \affiliation{Univ. Lyon, Universit\'e Lyon 1, CNRS/IN2P3, IP2I-Lyon, F-69622, Villeurbanne, France}
\author{H.J.~Birch} \affiliation{Department of Physics, University of Zurich, Winterthurerstrasse  190, 8057 Zurich, Switzerland}
\author{J.~Bl\'e} \affiliation{Univ. Grenoble Alpes, CNRS, Grenoble INP, LPSC-IN2P3, 38000 Grenoble, France}
\author{C.L. Chang} \affiliation{Argonne National Laboratory, 9700 S Cass Ave, Lemont, IL 60439, USA} \affiliation{Kavli Institute for Cosmological Physics, The University of Chicago, 5640 S Ellis Ave., Chicago, IL 60637} \affiliation{Department of Astronomy and Astrophysics, The University of Chicago, Eckhardt, 5640 S Ellis Ave., Chicago, IL 60637}
\author{Y.-Y.~Chang} \affiliation{University of California Berkeley, Department of Physics, 366 LeConte Hall 7300, Berkeley, CA 94720, USA}
\author{L.~Chaplinsky} \affiliation{University of Massachusetts, Amherst Center for Fundamental Interactions and Department of Physics, 101 Stockbridge Hall, 80 Campus Center Way, Amherst, MA 01003-9337 USA}
\author{G.~Cline} \affiliation{Lawrence Berkeley National Laboratory, 1 Cyclotron Rd., Berkeley, CA 94720, USA}
\author{A.~Cochard} \affiliation{Universit\'e Paris-Saclay, CNRS/IN2P3, IJCLab, 91405 Orsay, France}
\author{I.~Cojocari} \affiliation{Universit\'e Paris-Saclay, CNRS/IN2P3, IJCLab, 91405 Orsay, France}
\author{J.~Colas} \affiliation{Univ. Lyon, Universit\'e Lyon 1, CNRS/IN2P3, IP2I-Lyon, F-69622, Villeurbanne, France}
\author{M.~De~Jesus} \affiliation{Univ. Lyon, Universit\'e Lyon 1, CNRS/IN2P3, IP2I-Lyon, F-69622, Villeurbanne, France}
\author{P.~de~Marcillac} \affiliation{Universit\'e Paris-Saclay, CNRS/IN2P3, IJCLab, 91405 Orsay, France}
\author{K.~Dwinger} \affiliation{Department of Physics, University of Zurich, Winterthurerstrasse  190, 8057 Zurich, Switzerland}
\author{R.~Faure} \affiliation{Univ. Lyon, Universit\'e Lyon 1, CNRS/IN2P3, IP2I-Lyon, F-69622, Villeurbanne, France}
\author{S.~Fiorucci} \affiliation{Lawrence Berkeley National Laboratory, 1 Cyclotron Rd., Berkeley, CA 94720, USA}
\author{M.~Garcia-Sciveres} \affiliation{Lawrence Berkeley National Laboratory, 1 Cyclotron Rd., Berkeley, CA 94720, USA} \affiliation{International Center for Quantum-field Measurement Systems for Studies of the Universe and Particles (QUP,WPI), High Energy Accelerator Research Organization (KEK), Oho 1-1, Tsukuba, Ibaraki 305-0801, Japan}
\author{J.~Gascon} \affiliation{Univ. Lyon, Universit\'e Lyon 1, CNRS/IN2P3, IP2I-Lyon, F-69622, Villeurbanne, France}
\author{C.~Girard-Carillo} \affiliation{Univ. Grenoble Alpes, CNRS, Grenoble INP, LPSC-IN2P3, 38000 Grenoble, France}
\author{W.~Guo} \affiliation{Department of Mechanical Engineering, FAMU-FSU College of Engineering, Florida State University, 2525 Pottsdamer Street,  Tallahassee, FL 32310, USA} \affiliation{National High Magnetic Field Laboratory, 1800 E Paul Dirac Dr., Tallahassee, FL 32310, USA}
\author{L.~Haegel} \affiliation{Univ. Lyon, Universit\'e Lyon 1, CNRS/IN2P3, IP2I-Lyon, F-69622, Villeurbanne, France}
\author{S.J.~Haselschwardt} \affiliation{University of Michigan, Randall Laboratory of Physics, Ann Arbor, MI 48109-1040, USA}
\author{S.A.~Hertel} \affiliation{University of Massachusetts, Amherst Center for Fundamental Interactions and Department of Physics, 101 Stockbridge Hall, 80 Campus Center Way, Amherst, MA 01003-9337 USA}
\author{K.~Hunter} \affiliation{Texas A\&M University, Department of Physics and Astronomy, 4242 TAMU, 578 University Dr., College Station, TX 77843-4242, USA}
\author{L.~Juigne} \affiliation{Department of Physics, University of Zurich, Winterthurerstrasse  190, 8057 Zurich, Switzerland}
\author{A.~Juillard} \affiliation{Univ. Lyon, Universit\'e Lyon 1, CNRS/IN2P3, IP2I-Lyon, F-69622, Villeurbanne, France}
\author{A.~Kavner} \affiliation{Department of Physics, University of Zurich, Winterthurerstrasse  190, 8057 Zurich, Switzerland}
\author{J.~Lamblin} \affiliation{Univ. Grenoble Alpes, CNRS, Grenoble INP, LPSC-IN2P3, 38000 Grenoble, France}
\author{T.~Le-Bellec} \affiliation{Univ. Lyon, Universit\'e Lyon 1, CNRS/IN2P3, IP2I-Lyon, F-69622, Villeurbanne, France}
\author{X.~Li} \affiliation{Lawrence Berkeley National Laboratory, 1 Cyclotron Rd., Berkeley, CA 94720, USA}
\author{J.~Lin} \affiliation{University of California Berkeley, Department of Physics, 366 LeConte Hall 7300, Berkeley, CA 94720, USA} \affiliation{Lawrence Berkeley National Laboratory, 1 Cyclotron Rd., Berkeley, CA 94720, USA}
\author{R.~Mahapatra} \affiliation{Texas A\&M University, Department of Physics and Astronomy, 4242 TAMU, 578 University Dr., College Station, TX 77843-4242, USA}
\author{S.~Marnieros} \affiliation{Universit\'e Paris-Saclay, CNRS/IN2P3, IJCLab, 91405 Orsay, France}
\author{C.~Marrache} \affiliation{Universit\'e Paris-Saclay, CNRS/IN2P3, IJCLab, 91405 Orsay, France}
\author{N.~Martini} \affiliation{Univ. Lyon, Universit\'e Lyon 1, CNRS/IN2P3, IP2I-Lyon, F-69622, Villeurbanne, France}
\author{W.~Matava} \affiliation{University of California Berkeley, Department of Physics, 366 LeConte Hall 7300, Berkeley, CA 94720, USA} \affiliation{Lawrence Berkeley National Laboratory, 1 Cyclotron Rd., Berkeley, CA 94720, USA}
\author{D.N.~McKinsey} \affiliation{University of California Berkeley, Department of Physics, 366 LeConte Hall 7300, Berkeley, CA 94720, USA} \affiliation{Lawrence Berkeley National Laboratory, 1 Cyclotron Rd., Berkeley, CA 94720, USA}
\author{J.~Menu} \affiliation{Univ. Grenoble Alpes, CNRS, Grenoble INP, LPSC-IN2P3, 38000 Grenoble, France}
\author{K.~Moraa} \affiliation{Department of Physics, University of Zurich, Winterthurerstrasse  190, 8057 Zurich, Switzerland}
\author{V.~Novati} \affiliation{Univ. Grenoble Alpes, CNRS, Grenoble INP, LPSC-IN2P3, 38000 Grenoble, France}
\author{E.~Olivieri}\affiliation{Universit\'e Paris-Saclay, CNRS/IN2P3, IJCLab, 91405 Orsay, France}
\author{B.~Penning} \affiliation{Department of Physics, University of Zurich, Winterthurerstrasse  190, 8057 Zurich, Switzerland}
\author{M.~Platt} \affiliation{Texas A\&M University, Department of Physics and Astronomy, 4242 TAMU, 578 University Dr., College Station, TX 77843-4242, USA}
\author{M.~Pyle} \affiliation{University of California Berkeley, Department of Physics, 366 LeConte Hall 7300, Berkeley, CA 94720, USA} \affiliation{Lawrence Berkeley National Laboratory, 1 Cyclotron Rd., Berkeley, CA 94720, USA}
\author{D.~Poda}\affiliation{Universit\'e Paris-Saclay, CNRS/IN2P3, IJCLab, 91405 Orsay, France}
\author{Y.~Qi} \affiliation{Department of Mechanical Engineering, FAMU-FSU College of Engineering, Florida State University, 2525 Pottsdamer Street,  Tallahassee, FL 32310, USA} \affiliation{National High Magnetic Field Laboratory, 1800 E Paul Dirac Dr., Tallahassee, FL 32310, USA}
\author{M.~Reed} \affiliation{University of California Berkeley, Department of Physics, 366 LeConte Hall 7300, Berkeley, CA 94720, USA}
\author{R.K.~Romani} \thanks{Corresponding author: \href{mailto:rkromani@berkeley.edu}{rkromani@berkeley.edu}} \thanks{These authors contributed equally to this work} \thanks{now at: National Institute of Standards and Technology, Applied Physics Division, 325 Broadway, Boulder, CO 80305, USA}\affiliation{University of California Berkeley, Department of Physics, 366 LeConte Hall 7300, Berkeley, CA 94720, USA} 
\author{I.~Rydstrom} \affiliation{University of California Berkeley, Department of Physics, 366 LeConte Hall 7300, Berkeley, CA 94720, USA}
\author{B.~Sadoulet} \affiliation{University of California Berkeley, Department of Physics, 366 LeConte Hall 7300, Berkeley, CA 94720, USA}
\author{S.~Scorza} \affiliation{Univ. Grenoble Alpes, CNRS, Grenoble INP, LPSC-IN2P3, 38000 Grenoble, France}
\author{B.~Serfass} \affiliation{University of California Berkeley, Department of Physics, 366 LeConte Hall 7300, Berkeley, CA 94720, USA}
\author{P.~Sorensen} \affiliation{Lawrence Berkeley National Laboratory, 1 Cyclotron Rd., Berkeley, CA 94720, USA}
\author{S.~Steinfeld} \affiliation{University of California Berkeley, Department of Physics, 366 LeConte Hall 7300, Berkeley, CA 94720, USA} \affiliation{Lawrence Berkeley National Laboratory, 1 Cyclotron Rd., Berkeley, CA 94720, USA}
\author{H.~Su} \affiliation{University of Massachusetts, Amherst Center for Fundamental Interactions and Department of Physics, 101 Stockbridge Hall, 80 Campus Center Way, Amherst, MA 01003-9337 USA}
\author{A.~Suzuki} \affiliation{Lawrence Berkeley National Laboratory, 1 Cyclotron Rd., Berkeley, CA 94720, USA}
\author{R.L.~Vaughn~II} \affiliation{University of Massachusetts, Amherst Center for Fundamental Interactions and Department of Physics, 101 Stockbridge Hall, 80 Campus Center Way, Amherst, MA 01003-9337 USA}
\author{C.~Veihmeyer} \affiliation{University of California Berkeley, Department of Physics, 366 LeConte Hall 7300, Berkeley, CA 94720, USA}
\author{V.~Velan} \affiliation{Lawrence Berkeley National Laboratory, 1 Cyclotron Rd., Berkeley, CA 94720, USA}
\author{G. Wang} \affiliation{Argonne National Laboratory, 9700 S Cass Ave, Lemont, IL 60439, USA}
\author{P.~Vittaz} \affiliation{Univ. Lyon, Universit\'e Lyon 1, CNRS/IN2P3, IP2I-Lyon, F-69622, Villeurbanne, France}
\author{Y.~Wang} \affiliation{University of California Berkeley, Department of Physics, 366 LeConte Hall 7300, Berkeley, CA 94720, USA} \affiliation{Lawrence Berkeley National Laboratory, 1 Cyclotron Rd., Berkeley, CA 94720, USA}
\author{M.R.~Williams} \thanks{Corresponding author: \href{mailto:michaelwilliams@lbl.gov}{michaelwilliams@lbl.gov}} \thanks{These authors contributed equally to this work.} \affiliation{Lawrence Berkeley National Laboratory, 1 Cyclotron Rd., Berkeley, CA 94720, USA}
\author{J.~Wuko} \affiliation{University of Massachusetts, Amherst Center for Fundamental Interactions and Department of Physics, 101 Stockbridge Hall, 80 Campus Center Way, Amherst, MA 01003-9337 USA}

\collaboration{TESSERACT Collaboration}

\author{K. E. J. Myers}
\affiliation{University of California Berkeley, Department of Nuclear Engineering, Berkeley, CA 94720, USA} \affiliation{Lawrence Berkeley National Laboratory, 1 Cyclotron Rd., Berkeley, CA 94720, USA}

\author{L. Bernstein}
\affiliation{University of California Berkeley, Department of Nuclear Engineering, Berkeley, CA 94720, USA} \affiliation{Lawrence Berkeley National Laboratory, 1 Cyclotron Rd., Berkeley, CA 94720, USA}

\author{M. Potts}
\affiliation{Pacific Northwest National Lab, 902 Battelle Boulevard, Richland, WA 99354, USA}

\author{J. Orrell}
\affiliation{Pacific Northwest National Lab, 902 Battelle Boulevard, Richland, WA 99354, USA}

\date{\today}

\begin{abstract}
Solid state athermal phonon calorimeters used in the search for low mass dark matter or coherent neutrino-nucleus interactions have long observed a large excess of events below several hundred eV. The relaxation of damage created by the interaction of fast cosmic ray neutrons with the detector has been proposed as a source of these excess  events. By comparing neutron exposed detectors to control detectors, we report the first measurement of phonon bursts caused by damage created by fast neutrons. Differences in the spectral shape, the rate dependence on thermal history, and the observed spectral rate scaled to the neutron exposure between irradiated and control detectors suggest that our observed LEE backgrounds are not dominated by neutron damage-induced phonon bursts.
\end{abstract}

\maketitle


\section{Introduction}
Detecting rare, low-energy nuclear recoils, such as those created by low mass dark matter or coherent neutrino-nucleus interactions, remains one of the most challenging problems in modern detector development\cite{LEEReview,  TwoChannelLimits,cresstfuture,supercdms2026,ricochetcommission}. Cryogenic solid state detectors have achieved eV-scale thresholds capable of probing new parameter space for low-mass dark matter\cite{TwoChannelPaper, TwoChannelLimits, Run57Paper}, but have been limited by a high rate of poorly understood non-ionizing low background events \cite{adariEXCESSWorkshopDescriptions2022, LEEReview,TwoChannelLimits}.

While a number of mechanisms for this so-called ``Low Energy Excess'' (LEE) have been proposed and experimentally observed (including high energy particles causing scintillation in insulating materials around the detector\cite{HVeVR4}, stress associated with detector mounting\cite{astromFractureProcessesObserved2006, AnthonyPetersen2024}, and events originating in aluminum structures on the detector surface\cite{AnthonyPetersen2024, AlRelaxation, TwoChannelPaper}), the most problematic background remains unexplained, but seems to originate from within the bulk crystal forming the phonon detector's target\cite{Run57Paper}.

A mechanism for creating LEE in the detector substrate was recently proposed in Ref. \cite{DefectLEE}, where the authors posit that fast neutrons from e.g. cosmic rays secondaries (CRs) will interact with the detector substrate, creating damage in the crystal structure which relaxes over long time scales. The relaxation of such damage would create bursts of phonons originating in the substrate and scaling with its volume, consistent with observations\cite{Run57Paper}. Similar phenomena have also been described and studied through simulation focused on other contexts as shown in Refs.\cite{SiRadDamage1, SiRadDamage2}. This letter describes an attempt to probe this proposed damage hypothesis.

\section{Methods}

\begin{figure}
\includegraphics[width=1\columnwidth]{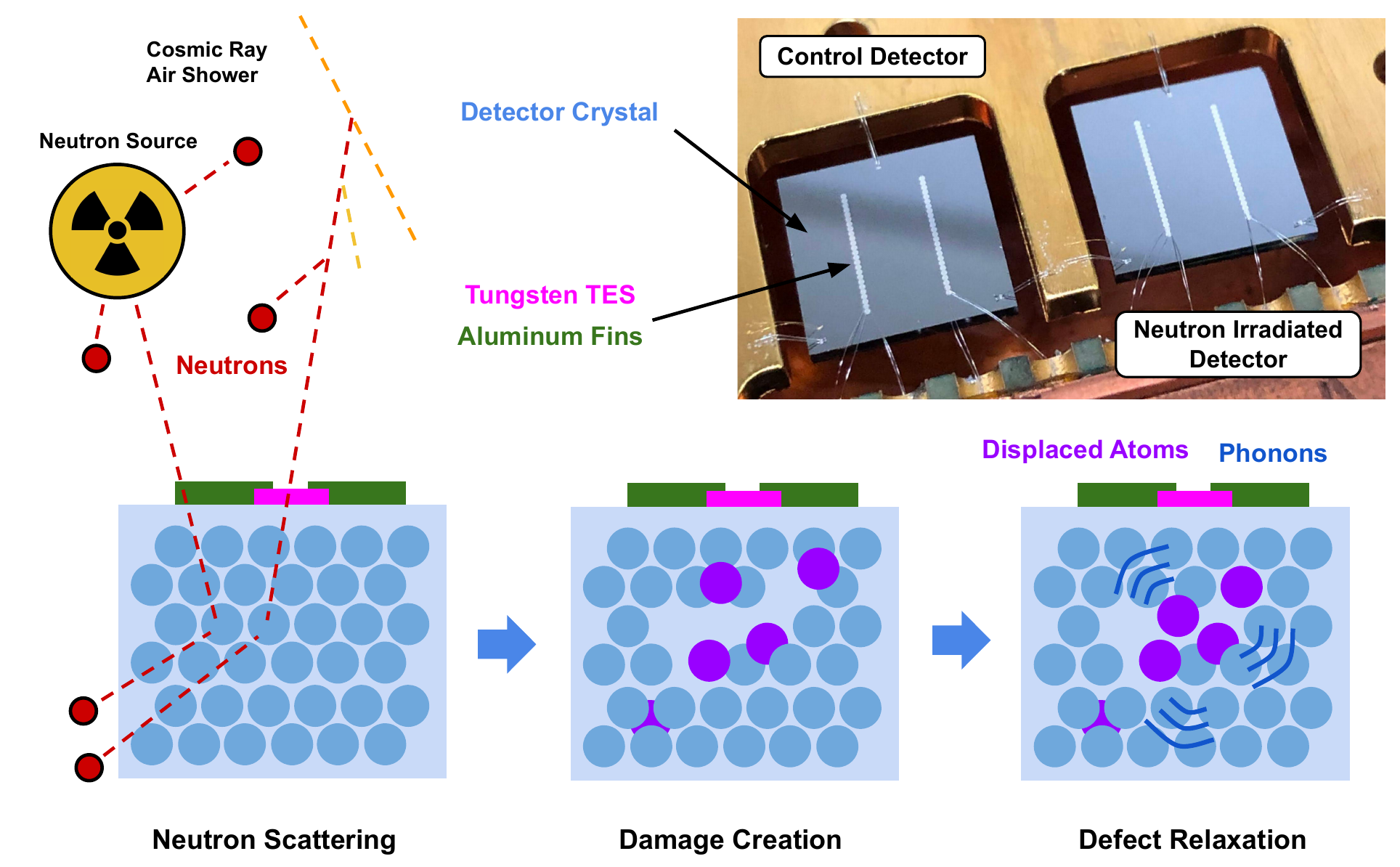}
\caption{\label{fig:diagram} Diagrams showing the proposed neutron damage and relaxation process. (Left) Fast neutrons recoil off individual atoms in the detector crystal. (Center Bottom) The energetic ``primary knock-on atom'' ricochets through the detector crystal, impacting additional atoms, creating extended defects. As atoms are displaced from the regular crystal lattice, this defect constitutes a metastable energetic state. (Right Bottom) After creation, the defect relaxes into a lower energy configuration, emitting athermal phonons. (Right Top) Photograph of the detectors, mounted next to each other in the cryogenic package. In each of the sub-diagrams, the tungsten TES films are marked in pink, while the aluminum films are marked in green.} 
\end{figure}

\subsection{Detectors}
To test if damage to the substrate lattice from fast neutrons can create LEE-like events, we compare LEE observations in two sets of two identically fabricated 1 mm thick by 1 cm$^2$ silicon athermal phonon detectors, one of which was exposed to a large fast neutron flux, while the other was kept unexposed as a control. These detectors use Transition Edge Sensors (TESs)\cite{irwinTransitionEdgeSensors2005} coupled to aluminum phonon collection fins in the common Quasiparticle-trap-assisted Electrothermal-feedback Transition-edge-sensor (QET)\cite{irwinQuasiparticleTrapAssisted1995} configuration. The control/irradiated pair was run simultaneously in a single optical cavity which was shielded from infrared radiation and electromagnetic interference and operated in a dilution refrigerator in a subbasement of the University of California, Berkeley physics department. The TESs were read out with single stage DC SQUID arrays.

The first set of detectors (set A) were an unmodified version of the detectors in Refs. \cite{TwoChannelPaper, TwoChannelLimits, Run57Paper}.  The second set of detectors (set B) were based on a modified version of the detectors in Refs. \cite{TwoChannelPaper, TwoChannelLimits, Run57Paper}, where the fin length was reduced by a factor of approximately two and the number of QETs and TES length were increased by approximately a factor of four. The resulting detector had a very similar aluminum coverage (and phonon dynamics) to the detectors in Refs. \cite{TwoChannelPaper, TwoChannelLimits, Run57Paper}, while the longer TESs increased the quiescent bias power and saturation energy of the detector by approximately four times. All detectors were manufactured on the same wafer as the 1 mm detector in Ref. \cite{Run57Paper}, and were suspended by wire bonds\cite{AnthonyPetersen2024}. We used two phonon readout channels per detector to separate ``shared'' (phonon-coupled) from ``singles'' (film-coupled) LEE\cite{TwoChannelPaper}.

\subsection{Neutron Exposures}

The irradiated (``rad'') detector in set A was exposed to both 2.45 MeV neutrons from a Deuterium-Deuterium (DD) generator for 5.8 days and 0-11 MeV neutrons from an Americium-Beryllium (AmBe) source for 24 hours, while the irradiated detector in set B was exposed only to DD neutrons for 4.8~days. Both detectors were exposed before mounting, and were located as close to the neutron sources as mechanically possible. The control detectors were not exposed to artificial neutron sources, but were unavoidably exposed to the same background neutron sources (e.g. cosmic ray secondaries) as our irradiated detectors. We expect that these artificial neutron exposures should negligibly activate our detectors (see appendix \ref{appendix:activation}), and that any resulting radioactive decays should contribute only high energy (keV+) events, far outside of our LEE region of interest. Methods for estimating the neutron-induced damage can be found in appendix \ref{appendix:exposures}, and the exposures in table \ref{table:neutron_irradiation}.

\subsection{Operation and Data Processing}

We cooled our detectors to millikelvin temperatures in a dilution refrigerator within several days of irradiation and began periodically recording 2 to 12 hour datasets 
to measure the LEE spectrum as a function of time since cooldown. Every 30 minutes of data-taking, we performed an IV sweep, a $\partial I/\partial V$ measurement at the TES operating point and a photon calibration to characterize the state of each channel and to account for drifts in response over time\cite{TwoChannelPaper, Run57Paper}. Approximately two months after the initial cooldown (Run 1) of the Set A detectors, we cooled them down a second time (Run 2) to monitor the evolution of their backgrounds over time.

After taking data, we characterized our detectors' Cross Spectral Density (CSD) and modeled their frequency dependent responsivity $dP/dI(\omega)$ by measuring their IV and $\partial I/\partial V$ responses. We determined the time domain current template for phonon pulses using the measured phonon response from photon calibrations\cite{TwoChannelPaper, Run57Paper}. Using these phonon templates and CSDs, we constructed a two-channel, one-amplitude optimal filter used as an offline software trigger and to measure the energy of phonon events\cite{TwoChannelLimits, Run57Paper}. 

In each detector, we observed intermittent mechanically excitable noise at \textasciitilde 200 Hz, which we interpret as a vibration mode of the wire bonds used to mechanically support the detectors. In some detectors, during periods of high environmental vibration excitation, we observed many low ($\lesssim$10 eV) energy phonon events occurring in phase with the noise. To remove periods with these events (which are unassociated with the higher energy neutron damage induced phonon bursts we are interested in studying), we measured the average current through our TESs in a roughly 200 ms long pre-pulse region, and remove from the analysis (``cut'') events for which this current exceeded a set threshold. Details can be found in appendix \ref{appendix:vibration_backgrounds}.

Additionally, we observed background events that were coincident between the two detectors, seemingly caused by scintillation of our PCB, electromagnetic interference, and possibly additional effects. To remove these events (which are also unassociated with the primary LEE effect we wish to study here), we cut events where there is an event in the other detector within 50 $\mu$s of the main event trigger (see appendix \ref{appendix:coincident_backgrounds}).

As in Refs. \cite{TwoChannelPaper, TwoChannelLimits, Run57Paper}, we applied a $\delta \chi^2$ based cut designed to discriminate ``singles'' from ``shared'' events. 
We also impose standard data quality cuts as in Refs. \cite{TwoChannelPaper, TwoChannelLimits, Run57Paper} which we further discuss in appendix \ref{appendix:quality_cuts}. We inject simulated waveforms at known rate and energy to estimate the passage fraction for each cut (see appendix \ref{appendix:cut_efficiency}).

\section{Results}

\begin{figure}
\includegraphics[width=1\columnwidth]{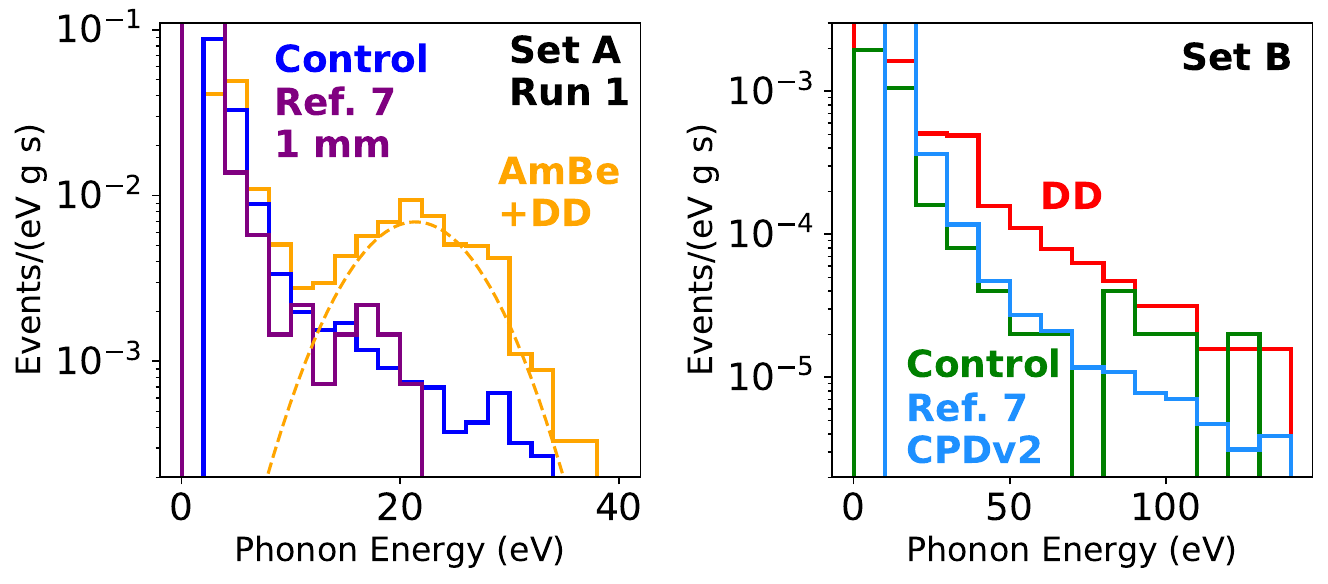}
\caption{\label{fig:spectrum} (Left) Set A spectra, measured approximately 6 hours after cooldown. Yellow corresponds to a detector irradiated with fast neutrons from AmBe and DD sources (see text), while blue corresponds to the control (unexposed) detector. For comparison, purple shows the spectrum on day 3.4 in the 1 mm detector in Ref. \cite{Run57Paper}. (Right) Spectra measured for the detectors in set B, measured approximately 1.1 days after cooldown. Red corresponds to the detector exposed to DD neutrons (see text), green is control, and light blue is the ``CPDv2'' detector spectrum shown in Ref. \cite{Run57Paper}.}
\end{figure}

In both detector sets, we independently observed a significant excess of events above roughly 10 eV in the irradiated detectors compared to the non-irradiated detectors (see Fig. \ref{fig:spectrum}). A larger excess was observed in set A, which we expect to have more neutron-induced damage (see table \ref{table:neutron_irradiation}).

In the set A detectors, the spectral shape of the excess events observed in the irradiated detector was significantly different than the control spectra. Specifically, we observe a Gaussian peak in the irradiated device spectrum centered around 20.1 $\pm$ 0.1 eV with a width of 4.4 $\pm$ 0.1 eV on top of the standard LEE background, with a tail towards higher energies. The measured detector resolution is significantly better than this width (see appendix \ref{appendix:resolutions}), indicating that the measured width is due to a physical process rather than a monoenergetic peak broadened by detector response. This peak is larger than the Frenkel pair formation energy \cite{defectenergy,Goedecker:PRL02:4foldDefectSi}and thus suggests the relaxation of more complex defect states.
In the set B irradiated detector, which has a higher dynamic range but roughly 3 times less exposure to neutrons, we do not see a clear peak, but instead a more uniformly enhanced background. 

\begin{figure}
\includegraphics[width=1\columnwidth]{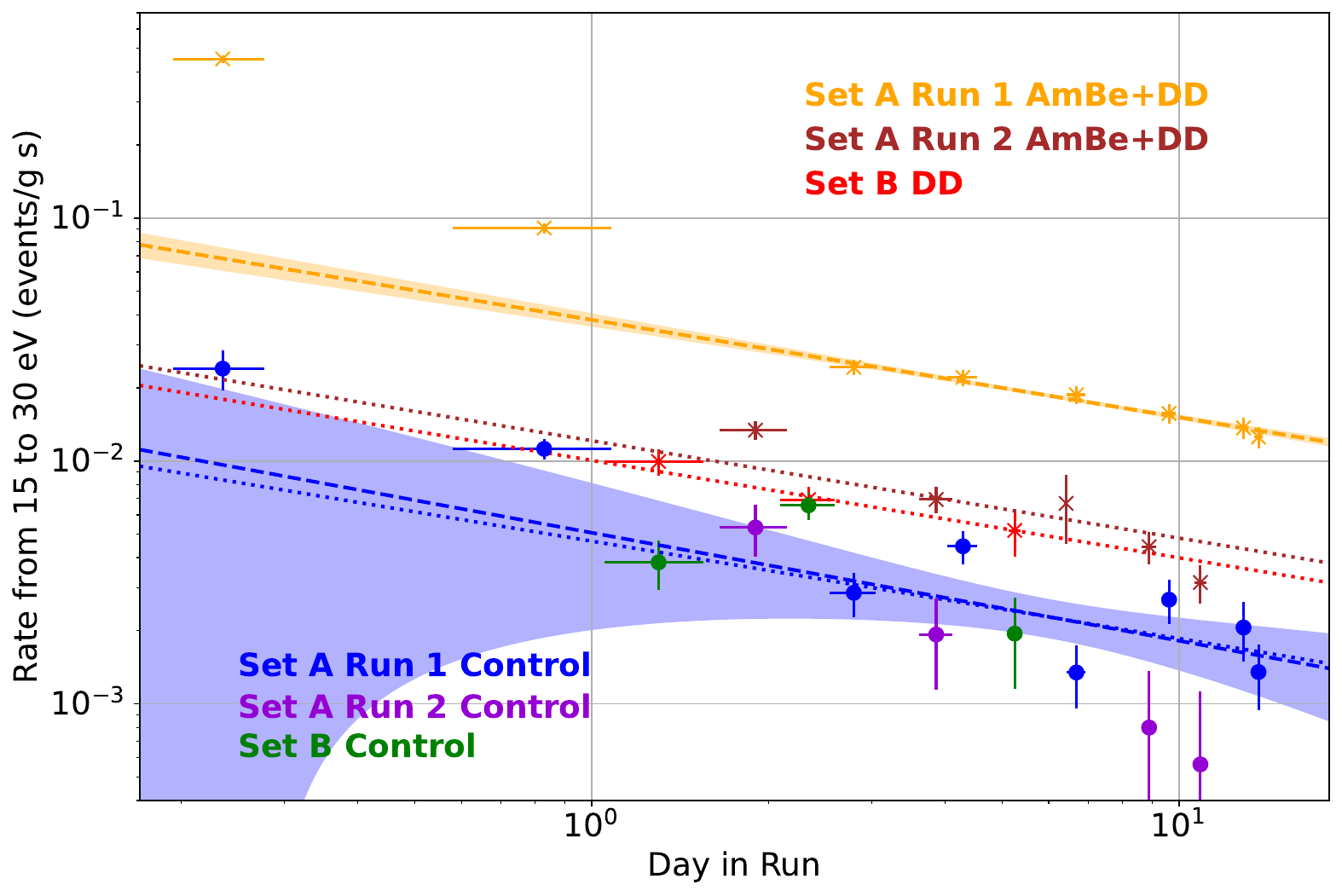}
\caption{\label{fig:time} Rate of events between 15 and 30 eV (where neutron-induced excess events dominate) observed in the detectors in sets A and B. Irradiated detectors are shown in bright colors with crosses, control detectors are shown in dark colors with dots. Dashed lines show fits to Eqn. \ref{eqn:pwr_law}, with 1 sigma uncertainties shaded. Dotted lines are fits with the power law exponent fixed (Eqn. \ref{eqn:pwr_law_fix}).}
\end{figure}

In both the irradiated and control detectors, we measured a decrease in shared LEE rate with time (see Fig. \ref{fig:time}). Considering this rate variation with time, the LEE rates observed in our control detectors were in good agreement with each other and with previously run detectors\cite{Run57Paper, TwoChannelPaper, TwoChannelLimits}, giving us some assurance that the excess backgrounds we observe in the irradiated detectors are due to neutron irradiation rather than some uncontrolled systematic variable.

To quantify the difference in rate between the irradiated and non-irradiated detectors, we measure the efficiency-corrected rate between 15 and 30 eV (where the neutron-induced excess appears to be largest), plotting the results in Fig. \ref{fig:time}. For detector set A, which we measured over the course of two runs (Set A Run 1, Set A Run 2), we fit this time series data to a power law
\begin{eqnarray}
\label{eqn:pwr_law}
    R_p(t) = \beta t^{-\alpha}
\end{eqnarray}
which we find describes the data reasonably well, except for the first two points, which we exclude from the fit. In Set A Run 1, we measure that the irradiated detector (dominated by excess backgrounds) and the control detector (observing backgrounds consistent with previous observations\cite{Run57Paper}) and see statistically consistent $\alpha$ exponents ($\alpha_{rad} = 0.40 \pm 0.03$, $\alpha_{cont} = 0.4 \pm 0.3$). 

\begin{figure}
\includegraphics[width=1\columnwidth]{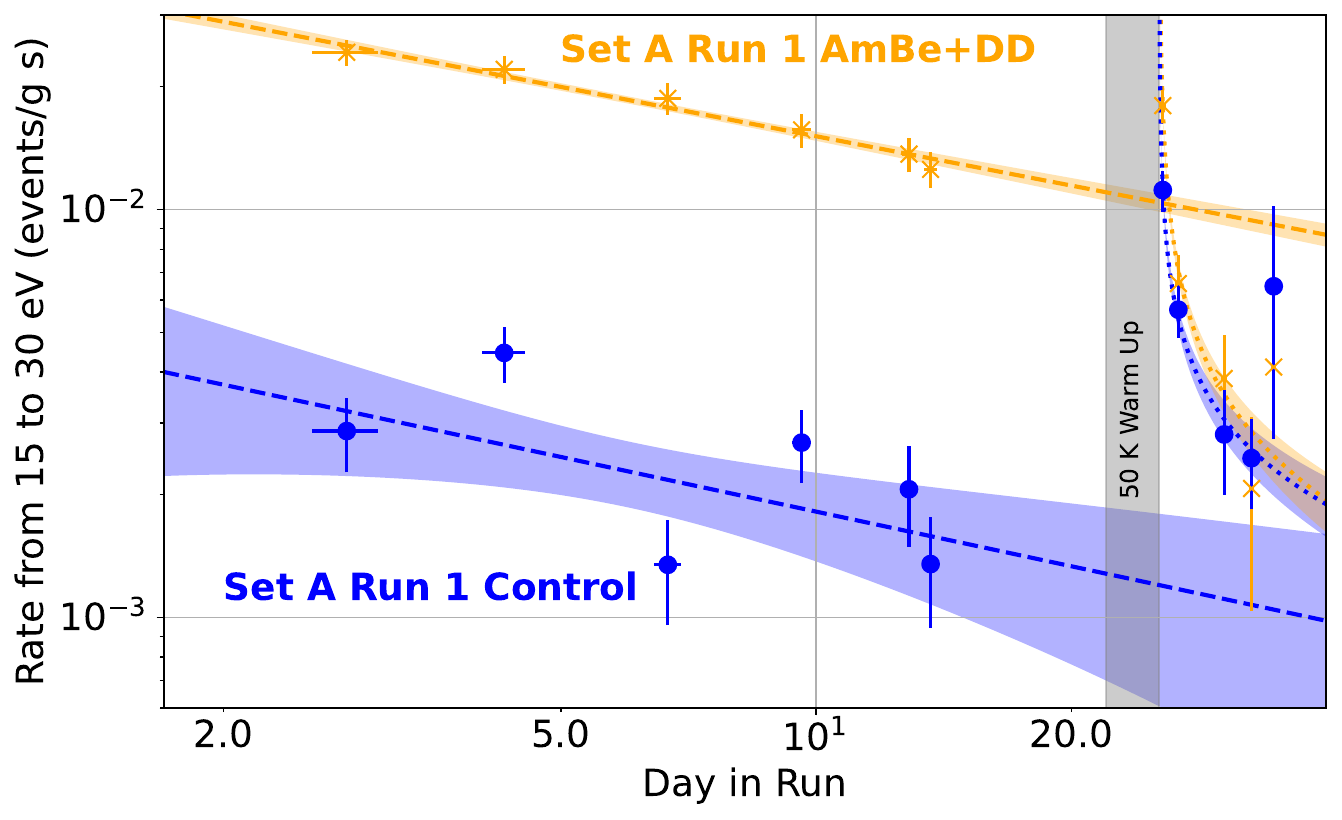}
\caption{\label{fig:warmup} Rate of 15-30 eV events in the set A detectors, showing the change in observed rate after warming up to 50 K for three days. Dashed lines show the same fits as in Fig. \ref{fig:time}, while dotted lines show additional power law fits (Eqn. \ref{eqn:pwr_law_warmup}). Note that the warm up period significantly reduces the observed background in the irradiated device.}
\end{figure}

Approximately one month into Run 1 of the set A detectors, we warmed up to $\sim$50 K for three days before recooling to base temperature and resuming measurements. Consistent with previous LEE measurements\cite{angloherLatestObservationsLow2022}, we found that the observed LEE rate was enhanced before relaxing (see Fig. \ref{fig:warmup}). After this warm up, we fit to a power law model
\begin{eqnarray}
\label{eqn:pwr_law_warmup}
    R_{\mathrm{wu}}(t) = \beta (t - t_{\mathrm{wu}})^{-\alpha}
\end{eqnarray}
where $t_{\mathrm{wu}}$ is the fixed time of the end of the warm up. We measure essentially consistent rates and time dependencies between the irradiated and control devices, suggesting that the warm up succeeded in reducing the relative impact of neutron damage to the phonon burst rate. 

Additionally, we re-measured the detectors in set A approximately two months after initially cooling them down (Set A Run 2), finding that while the time-corrected LEE rate in the control detector was unchanged, the rate of phonon bursts in the irradiated detector had decreased, although it remained higher than the control detector. To quantify this change in rate, we fix the power law parameter to the measured $a_{rad}$ value, and fit only the rate constant $b$.
\begin{eqnarray}
\label{eqn:pwr_law_fix}
    R_f(t) = \beta t^{-\alpha_{\mathrm{rad}}}
\end{eqnarray}
We measure that $\beta_{A1,\mathrm{rad}}/\beta_{A1,\mathrm{control}} = 8 \pm 1$, $\beta_{A2,\mathrm{rad}}/\beta_{A1,\mathrm{control}} = 2.6 \pm 0.6$, and $\beta_{B,\mathrm{rad}}/\beta_{A1,\mathrm{control}} = 2.1 \pm 0.3$. While the relaxation of these neutron induced defects is apparently non-trivially temperature dependent, we can use the two set A runs to estimate the lifetime of neutron-induced defects in the substrate at room temperature over the months timescale as
\begin{eqnarray}
    \tau \approx \frac{\Delta t}{\ln \big(\beta_{A1,\mathrm{rad}} / \beta_{A2,\mathrm{rad}} \big)} \approx 75.06 \pm 0.03 \mathrm{\ days} 
\end{eqnarray}
where $\Delta t$ is the time between runs 1 and 2.

\section{Discussion}

We have unambiguously observed low energy phonon bursts created by the relaxation of fast neutron-induced damage in our detector crystals. However, such events might only contribute subdominantly to the LEE spectrum in detectors seeing only background neutron radiation.

To estimate the contribution of cosmic ray neutron induced damage to the observed spectrum of LEE events, we model the damage created by our neutron sources and compare it to the expected damage from cosmic rays. Here, we model the cosmic ray neutron spectrum using EXPACS\cite{EXPACS1, EXPACS2, EXPACS3, EXPACS4}, and use both a Geant4 + Lindhard yield\cite{Lindhard1961} model and a tabulated Non-Ionizing Energy Loss (NIEL) model for neutrons incident on silicon \cite{NIELSi} to estimate the damage induced by neutrons in the crystal. We assume that the dominant damage created by cosmic rays and our artificial neutron sources are both above the energy threshold where neutron events create many similar sub-cascades and defects\cite{DefectLEE, SiRadDamage1, SiRadDamage2}, meaning the amount of damage produced (and the number of relaxation events we expect) is proportional to the total non-ionizing energy deposited in the crystal by neutrons.

To make an equal-footed comparison of the detectors exposed continuously to cosmic rays with those exposed to artificial neutron sources over a known period of time, we must select a period of time over which to integrate the cosmic ray flux. Here, we choose two possibly relevant time scales. First, the cosmic ray induced damage may relax away slowly compared to the age of the crystal, i.e. the damage is integrated over the entire crystal lifetime (see table \ref{tab:cr_history}). Alternatively, the neutron damage may relax over time scales comparable to our experiments, i.e. the roughly 75 day timescale we measure using the evolution of the phonon burst rate with detectors in set A. For completeness, we present results using both assumptions.

We calculate that the neutron damage from irradiation is 3 to 4 orders of magnitude larger than from cosmic ray neutrons (over either a 75 day damage lifetime or the entire crystal lifetime). However, the irradiated and control spectra differ by at most an order of magnitude. We summarize our results in table \ref{table:neutron_irradiation} and our methodology in appendix \ref{appendix:exposures}.

\begin{table*}[t]
\caption{\label{table:neutron_irradiation} Fast neutron energy deposited in each detector from our DD generator exposures, our AmBe exposures, and ambient cosmic rays. The 75 day cosmic ray exposure assumes the detector is on the ground at sea level. ``All'' indicates that each detector should experience an essentially identical cosmic ray exposure, and time spent on the ground and time on airplanes at high altitude since the wafer was created.}
\centering
\begin{center}
\begin{tabular}{ |c|c|c|c|c|c| } 
 \hline
 Detector & Irradiation Source & Irradiation Time & Time until Fridge Cold & Integrated Energy & Integrated Energy \\ 
  &  &  &  & Geant4 + Lindhard & NIEL \\
 \hline
 Rad, A & DD & 140 hr & 17.0 d & 318 GeV & 312 GeV \\
 Rad, A & AmBe & 24 hr & 9.75 d & 312 GeV & 469 GeV \\
 Rad, A & DD + AmBe & & & 630 GeV & 718 GeV \\
 \hline
 Rad, B & DD & 114.5 hr & 3.5 d & 256 GeV & 255 GeV \\
 \hline
 All & CR Neutrons & 75 d & & 0.052 GeV & 0.041 GeV \\
 All & CR Neutrons & All & & 1.15 GeV & 0.89 GeV \\
 \hline
\end{tabular}
\end{center}
\end{table*}

The relaxation of AmBe and DD neutron-induced damage does not create phonon bursts which are identical to the LEE. To highlight the main differences:
\begin{itemize}
    \item The spectral shapes of LEE in the control detectors and backgrounds in the irradiated detectors differ (although to varying degrees).
    \item The effects of neutron damage appear to relax away quickly at 50 K, bringing the measured rates and spectra in the irradiated and control detectors into much closer agreement.
    \item The rate of observed LEE after irradiation (compared to the control) is not as high as would be expected from scaling neutron damage models from cosmic rays to the irradiation dose.
\end{itemize}

\section{Conclusion}

Our work has demonstrated a new neutron-induced background for athermal phonon calorimeters: exposing low threshold silicon detectors to high neutron fluxes will create damage which causes low energy phonon bursts even long after exposure. Collaborations using low threshold nuclear recoil detectors which are exposed to large neutron fluxes (e.g. calibration sources) should therefore carefully consider whether neutron-induced damage will impact their science program.

More broadly, we observe significant tension with the hypothesis\cite{DefectLEE} that cosmic ray neutron-induced damage relaxation dominates the LEE spectrum. Moving forward, we are faced with three possibilities.

First, while phonon-coupled LEE may be primarily caused by the relaxation of cosmic ray induced defects, some key subtlety of the damage mechanism or our experiment might obscure the underlying similarity between artificially induced damage-relaxation events and natural LEE. For example, the higher energy neutrons present in cosmic rays could create significantly different damage complexes which emit phonon bursts with a higher rate and broader spectrum than our neutron sources. 

Second, natural LEE could be created predominantly by cosmic rays, but the annealing of this damage may have some non-trivial temperature or time dependence that could change the relationship between LEE rate and neutron damage. 

For both points, some experimental evidence supports the general hypothesis that cosmic rays create the LEE: both NUCLEUS\cite{bossio2025nucleus} and CDMS CPD detectors\cite{GermondThesis} have observed that the LEE rate drops over time while detectors are kept underground, with NUCLEUS observing that this rate increases again when the detectors are returned to the surface, before again decreasing with time after moving underground\cite{bossio2025nucleus}. These observations seem to suggest a slowly relaxing cosmic ray origin for the LEE.

Third, the dominant mechanism for LEE creation could be unrelated to cosmic rays. For example, defects created during crystal growth might later relax and create LEE phonon bursts. However, this proposal is difficult to reconcile with our previous observations that the volume-normalized LEE rate is consistent in different silicon detectors after correcting for time since cool down\cite{Run57Paper}. Naively, the defect density would be expected to vary between different silicon crystals, which would presumably lead to LEE rates which vary more than we observe.

In summary, we have observed that damage created by fast neutrons interacting with silicon crystals creates low energy phonon bursts emitted days to months after exposure. Several lines of evidence, including the spectral shape of the background, the expected rate when scaling from the expected cosmic ray exposure, and the rate after warm up to 50 K lead us to conclude that our observed LEE is not dominated by the relaxation of damage created by cosmic ray neutrons. Understanding and resolving the longstanding problem of phonon-coupled LEE will require further work, more precisely testing the cosmic ray hypothesis as well as the role of other defects in bulk crystals.

\section{Acknowledgments}

We thank Antoine Jay and Kai Nordlund for their illuminating discussions and comments.  

This work was supported in part by DOE Grants DE-SC0019319, DE-SC0025523 and DOE Quantum Information Science Enabled Discovery (QuantISED) for High Energy Physics (KA2401032). This material is based upon work supported by the National Science Foundation Graduate Research Fellowship under Grant No. DGE 1106400, the Department of Energy National Nuclear Security Administration through the Nuclear Science and Security Consortium under Award Number(s) DE-NA0003180 and/or DE-NA0000979, and by an appointment to the Intelligence Community Postdoctoral Research Fellowship Program at the National Institute of Standards and Technology administered by Oak Ridge Institute for Science and Education (ORISE) through an interagency agreement between the U.S. Department of Energy and the Office of the Director of National Intelligence (ODNI). Work at Lawrence Berkeley National Laboratory was supported by the U.S. DOE, Office of High Energy Physics, under Contract No. DEAC02-05CH11231. Work at Argonne is supported by the U.S. DOE, Office of High Energy Physics, under Contract No. DE-AC02-06CH11357. W.G. and Y.Q. acknowledge the support by the National High Magnetic Field Laboratory at Florida State University, which is supported by the National Science Foundation Cooperative Agreement No. DMR-2128556 and the state of Florida. Pacific Northwest National Laboratory is operated by Battelle Memorial Institute under contract No. DE-AC05-76RL01830 for the US Department of Energy (DOE) and contributions to this work were supported by the US DOE Oﬃce of High Energy Physics’s Cosmic Frontier program. This work benefited from state support managed by the French National Research Agency under the France 2030 program, reference ANR-24-RRII-0001.

The authors have no conflicts to disclose. The data that support the findings of this study are available from the corresponding author upon reasonable request.

\bibliography{aipsamp_.bib}

\appendix

\section{Neutron Activation}
\label{appendix:activation}

While we take care to expose our detectors to neutrons before packaging them to avoid activating their packaging, during our neutron irradiation we could conceivably be activating the detectors themselves. To estimate the possible importance of this potential activation, we multiply the total neutron fluence through the detector for the 1 day AmBe exposure of the set A detector by a conservatively chosen benchmark cross section given by the ENDF database between 1 and 10 MeV for three processes: (n, $\gamma$), (n, p), and (n, $\alpha$) (see table \ref{tab:activation})

Even with conservative assumptions about the timing of the neutron flux and the cross section of a given process, we see that every activation product considered here is either too low in activity to plausibly explain our low energy phonon burst observations, or too low in half life (i.e., the activated atoms would have decayed to a stable state by the time our detectors are cold and taking data). Combined with the fact that $\beta$s would be expected to create very high energy (keV+) events in our detector, we conclude that neutron activation contributes negligibly to our LEE observations. 

\begin{table*}[h]
\caption{\label{tab:activation} Summary of the decay of radioactive isotopes created by neutron activation following the AmBe neutron irradiation for the Run 62 devices. The ``Estimated $\sigma$" column gives an estimate of the maximum neutron cross section for a given reaction in the MeV to 4 to 10 MeV range, at which energies our sources emit neutrons. The activity column assumes that the detector instantaneously receives the radiation dose, and gives the activity immediately after exposure (i.e. before any decay). Both assumptions are conservative.}
\label{tab:activationtable}
\begin{tabular}{|c|c|c|c|c|}
\hline
\textrm{Reaction} & Product Half Life & Product Decay Channel & Estimated $\sigma$ & Estimated Activated Activity\\
\hline
$^{28}$Si + n $\rightarrow$ $^{28}$Al + p & 2.2 min & 4.6 MeV $\beta$ & 0.3 b & 7.6 kBq \\
$^{28}$Al $\rightarrow$ $^{28}$Si + $\beta$ & Stable & & & \\
\hline
$^{28}$Si + n $\rightarrow$ $^{25}$Mg + $\alpha$ & Stable & & &  \\
\hline
$^{29}$Si + n $\rightarrow$ $^{29}$Al + p & 6.7 min & 3.7 MeV $\beta$ & 0.1 b & 43 Bq \\
$^{29}$Al $\rightarrow$ $^{29}$Si + $\beta$ & Stable & & & \\
\hline
$^{29}$Si + n $\rightarrow$ $^{26}$Mg + $\alpha$ & Stable & & &  \\
\hline
$^{30}$Si + n $\rightarrow$ $^{30}$Al + p & 3.6 s & 8.6 MeV $\beta$ & 50 mb & 1.6 kBq \\
$^{30}$Al $\rightarrow$ $^{30}$Si + $\beta$ & Stable & & & \\
\hline
$^{30}$Si + n $\rightarrow$ $^{27}$Mg + $\alpha$ & 9.5 min & 2.6 MeV $\beta$ & 0.1 b & 20 Bq \\
$^{27}$Mg + $\beta$ $\rightarrow$ $^{27}$Al & Stable & & & \\
\hline
$^{30}$Si + n $\rightarrow$ $^{31}$Si + $\gamma$ & 2.6 h & 1.5 MeV $\beta$ & 1 mb & 12 mBq \\
$^{31}$Si $\rightarrow$ $^{31}$P + $\beta$ & Stable & & &  \\
\hline
\hline
$^{27}$Al + n $\rightarrow$ $^{27}$Mg + p & 9.5 min & 2.6 MeV $\beta$ & 0.1 b & 166 mBq \\
$^{27}$Mg $\rightarrow$ $^{27}$Al + $\beta$ & Stable & & &  \\
\hline
$^{27}$Al + n $\rightarrow$ $^{24}$Na + $\alpha$ & 15.0 h & 5.5 MeV $\beta$ & 0.1 b & 1.7 mBq \\
$^{24}$Na $\rightarrow$ $^{24}$Mg + $\beta$ & Stable & & & \\
\hline
\hline
$^{182}$W + n $\rightarrow$ $^{182}$Ta + p & 114 d & 1.8 MeV $\beta$ & 30 mb & 61 pBq \\
$^{182}$Ta $\rightarrow$ $^{182}$W + $\beta$ & Stable & & & \\
\hline
$^{182}$W + n $\rightarrow$ $^{179}$Hf + $\alpha$ & Stable & & & \\
\hline
$^{183}$W + n $\rightarrow$ $^{183}$Ta + p & 5.1 d & 1.1 MeV $\beta$ & 0.15 b & 3.7 nBq \\
$^{183}$Ta $\rightarrow$ $^{183}$W + $\beta$ & Stable & & & \\
\hline
$^{183}$W + n $\rightarrow$ $^{180}$Hf + $\alpha$ & Stable & & & \\
\hline
$^{184}$W + n $\rightarrow$ $^{185}$W & 75.1 d & 0.43 MeV $\beta$ & 0.1 b & 0.4 nBq \\
$^{185}$W $\rightarrow$ $^{185}$Re + $\beta$ & Stable & & & \\
\hline
$^{184}$W + n $\rightarrow$ $^{184}$Ta + p & 8.7 hr & 2.9 MeV $\beta$ & 10 mb & 8.3 nBq \\
$^{184}$Ta $\rightarrow$ $^{184}$W + $\beta$ & Stable & & & \\
\hline
$^{184}$W + n $\rightarrow$ $^{181}$Hf + $\alpha$ & 42.4 d & 1.0 MeV $\beta$ & 2 mb & 14 pBq \\
$^{181}$Hf $\rightarrow$ $^{181}$Ta + $\beta$ & Stable & & & \\
\hline
$^{186}$W + n $\rightarrow$ $^{187}$W & 23.8 h & 1.3 MeV $\beta$ & 40 mb & 142 pBq \\
$^{187}$W $\rightarrow$ $^{187}$Re + $\beta$ & $4.4\times 10^{10}$ yr & 3 keV $\beta$ & & \\
\hline
$^{186}$W + n $\rightarrow$ $^{186}$Ta + p & 10.5 min & 3.9 MeV $\beta$ & 10 mb & 340 nBq \\
$^{186}$Ta $\rightarrow$ $^{186}$W + $\beta$ & Stable & & & \\
\hline
$^{186}$W + n $\rightarrow$ $^{183}$Hf + $\alpha$ & 1.1 hr & 2.0 MeV $\beta$ & 2 mb & 11 nBq \\
$^{183}$Hf $\rightarrow$ $^{183}$Ta + $\beta$ & 5.1 d & 1.1 MeV $\beta$ & & \\
$^{183}$Ta $\rightarrow$ $^{183}$W + $\beta$ & Stable & & & \\
\hline
\end{tabular}
\end{table*}

\section{Neutron Exposures}
\label{appendix:exposures}

\begin{figure}
\includegraphics[width=1\columnwidth]{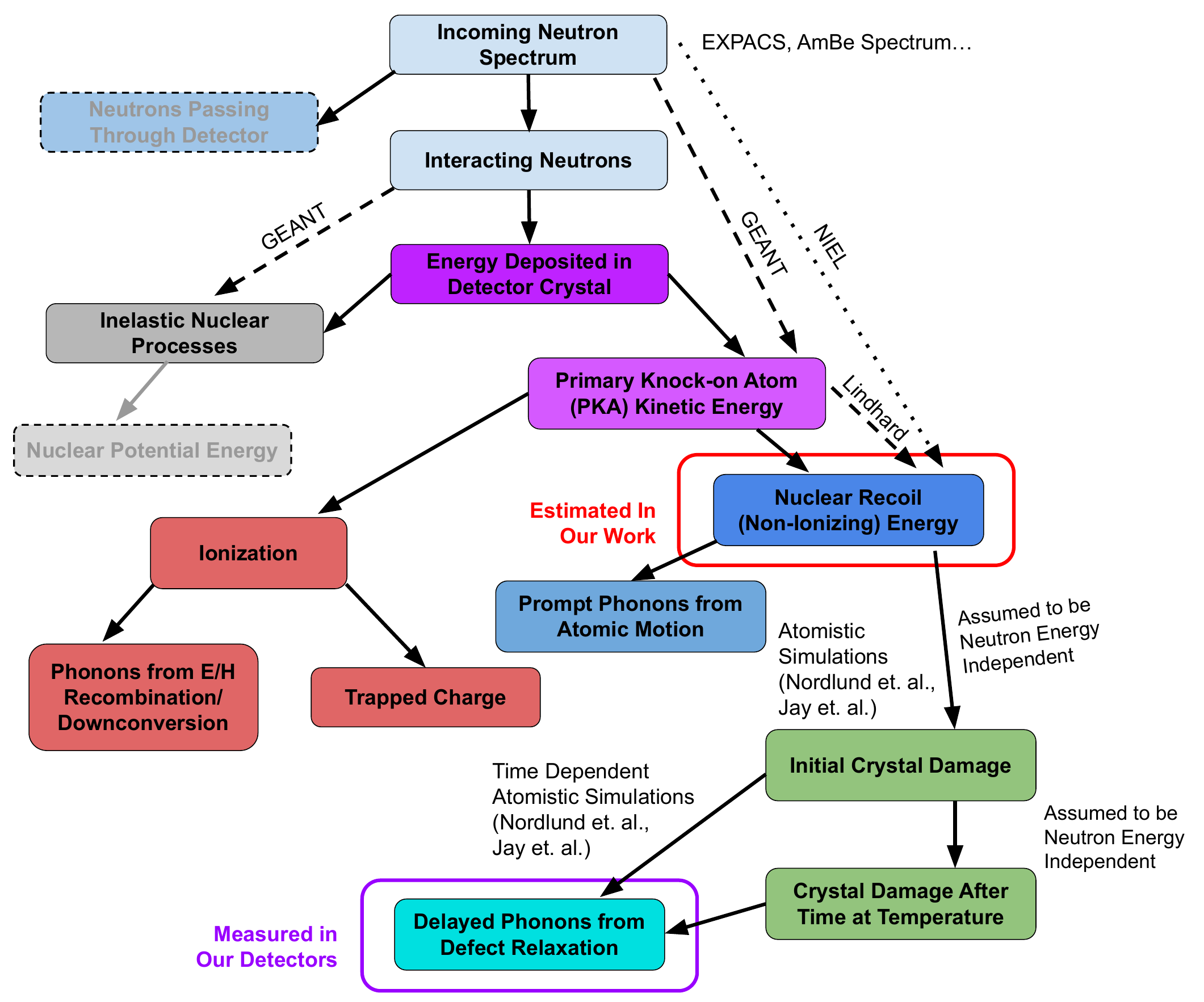}
\caption{\label{fig:sim_tree} Diagram showing the channels energy can partition into following a neutron interaction in the a detector crystal. We determine the spectrum of incoming cosmic ray neutrons using EXPACS\cite{EXPACS1, EXPACS2, EXPACS3, EXPACS4}, and take the AmBe spectrum from Ref. \cite{AmBeStudy}. Here, we estimate the amount of (non-ionizing) nuclear recoil energy as a proxy for the total amount of damage done to the crystal, which we assume is related to the rate of delayed phonon bursts from damage relaxation, i.e. LEE-type events. We make this estimate using two techniques (see text): a NIEL based model which directly estimates the nuclear recoil energy from the incoming neutron spectrum, and a model that uses Geant4 to simulate the energy deposited in the crystal by incoming neutrons and a Lindhard yield\cite{Lindhard1961} model to estimate the nuclear recoil energy (as opposed to ionizing energy) created in these events.}
\end{figure}

In this paper, we attempt to compare the damage created in our silicon detectors by cosmic ray neutrons to the damage created by our artificial neutron sources (AmBe and DD sources) to determine whether or not fast neutron induced damage can plausibly explain our LEE observations. To clarify this concept, we separate the damage creation process into three parts. First, we must understand the spectrum of neutrons which could interact with our detector; second, we attempt to characterize the amount of damage-inducing energy deposited in the detector by these neutrons; and third, we must consider what low energy phonon bursts we expect to be produced by this damage. See Fig. \ref{fig:sim_tree} for a graphical presentation of the channels neutrons can deposit energy into.

Estimating the integrated neutron spectrum incident on our detectors is relatively straightforward, with the largest uncertainty being the comic ray exposure time. To calculate the amount of damage-inducing energy that these neutrons deposit in our detectors, we use two techniques. In the first technique, we simulate neutrons incident on the detector using Geant4, counting the number of interactions and the energy deposited in these interactions (we do not count energy deposited by secondary gammas, electrons, and positrons, which only deposit energy electromagnetically and which cannot create significant damage). We then employ a Lindhard yield factor which models the fraction of energy which is deposited in non-ionizing (i.e. potentially damage inducing) channels. The second technique uses the standard Non-Ionizing Energy Loss (NIEL) approach in silicon, which is commonly used in collider physics~\cite{NIELSi}. 

Here, we make the simplifying assumption that the rate of phonon bursts from this damage is proportional to the total energy creating the damage. Simulations \cite{SiRadDamage1, SiRadDamage2, DefectLEE} seem to indicate that above a relatively low threshold ($\sim$ 1-10 keV silicon Primary Knock-on Atom, PKA, or $\sim$ 300 keV incident neutron) the damage induced by neutrons splits up into many sub-cascades, producing similar defect clusters with the number of clusters simply proportional to the energy initially deposited by the incident neutron. Further studies, both experimental and simulation-based, could focus on the potential for variations in the spectra created by neutrons at various energies.

The total exposure to neutrons from different sources (our DD generator, the AmBe source, and cosmic ray secondaries) is summarized in table \ref{table:neutron_irradiation}. We expand on the estimation of these quantities below.

\subsection{Geant4 and Damage-Inducing Yield Simulations}
\label{appendix:g4lindhard}

We use Geant4 (version v11.3.2) simulations with the QGSP\_BERT\_HP physics list to estimate both the probability of detector interaction and the nuclear recoil energy distribution. To increase statistics, we create a simulation environment in which we send a collimated beam of neutrons through a stack of fifty 1 mm thick by 1 cm$^2$ silicon detectors. For DD neutrons, we assume that they are monoenergetic, with an energy of 2.45 MeV, while in the case of AmBe neutrons, we draw their energies from a spectrum taken from Ref. \cite{AmBeStudy}. The spectrum of cosmic ray neutrons is too steeply sloped toward low energies to simulate the full distribution with adequate statistics when sampling directly from the EXPACS model. Instead, we sample 24 logarithmically spaced energies between 10 keV and 10 GeV, and use these to create a model we can interpolate and weight by the EXPACS spectrum as described below. From these simulations, we extract the fraction of neutrons which interact with a detector $f_i(E_n)$, and the average energy they deposit after interaction $E_{dep}(E_n)$.

However, part of the energy deposited by an incident neutron appears as ionization, with the remainder going to the non-ionizing channel. Here, we assume that this non-ionizing energy (or a broadly energy independent fraction of this energy) is what creates the damage to the silicon crystal. We use the parametrization of the Lindhard model\cite{Lindhard1961} given in Ref. \cite{NIELSi} for silicon, and note that different parameterizations of this model give significant deviations in the yield, especially at high energies. Here, we model the non-ionizing yield as
\begin{eqnarray}
    f_{NI} = \frac{1}{1 + k_d g(\epsilon_d)}
\end{eqnarray}
where
\begin{eqnarray}
    k_d = 0.1462 \\
    g(\epsilon_d) = \epsilon_d + 0.40244\epsilon_d^{3/4} + 3.4008\epsilon_d^{1/6} \\
    \epsilon_d = 2.147 \times 10^{-5} \mathrm{eV}^{-1} E_{dep} \approx \frac{E_{dep}}{46.58 \mathrm{keV}}
\end{eqnarray}
where $E_{dep}$ is the total energy deposited in the silicon by the incident neutron, following the values in Ref. \cite{NIELSi}. Note that at high energies, $g(\epsilon_d) \rightarrow \epsilon_d$, so the total energy deposited in the non-ionizing (i.e. damage inducing) channel goes as
\begin{eqnarray}
    \lim_{E_{dep} \to \infty} E_{dep} f_{NI} = \frac{E_{dep}}{1 + k_d E_{dep} / (46.58 \mathrm{\ keV})} \approx 319 \mathrm{\ keV}
\end{eqnarray}
i.e., within this model, no particle can deposit more than 319 keV of energy in a damage-creating channel.

To arrive at the total ``damage-inducing'' energy imparted on our detectors by incident neutrons, we find the average energy deposited per interaction $E_{dep}$ from neutrons with an initial kinetic energy of $E_n$, and evaluate the Lindhard non-ionizing yield model $f_{NI}(E_{dep})$ at the energy of interest (or weighted average energy, in the case of the AmBe spectrum). The total damage inducing energy per incident neutron deposited in the potentially damage inducing channel is therefore
\begin{equation}
    E_D(E_n) = f_i(E_n) f_{NI}(E_{dep}(E_n))E_{dep}(E_n)
\end{equation}
\begin{equation}
    E_D(E_n) = \int  f_{NI}(E_{dep}) E_{dep} \frac{df_{i}}{dE_{dep}}(E_{dep},E_n) dE_{dep}
\end{equation}

\subsection{Non-Ionizing Energy Loss (NIEL)}
We use the SR-NIEL dose calculator to convert our estimated neutron flux from various sources to a total amount of neutron damage in the lattice~\cite{sr-niel_history}. This method of estimating neutron damage to the lattice is expressed by the displacement KERMA function

\begin{equation}
D(E)=\sum_k\sigma_k(E)\int f_k(E,E_R)\,P_k(E_R)\,\mathrm{d}E_R.
\end{equation}
Here $\sum_k$ implies that we are summing over various interaction types (elastic,inelastic,etc.), $\sigma_k(E)$ is the cross section for a given reaction for a given incoming neutron kinetic energy, $f_k(E,E_R)\,\mathrm{d}E_R$ is the probability distribution of recoil energies for an interaction, and $P_k(E_R)$ is the partition energy of the recoil nucleus, i.e. how much of the total energy goes into potential displacements, typically evaluated using the Lindhard partition function~\cite{NIELSi}. This function has been measured and is readily available for use in calculations\cite{sr-niel_history}. We can then estimate the energy density that goes into displacements for a given fluence of neutrons:

\begin{equation}
E_{dis}=N_A \int_{E_{min}}D(E)\phi(E)dE
\end{equation}
where $N_A$ the number of atoms per cm$^3$ for a given material, $E_{min}$ is the minimum energy needed for an incoming neutron to cause damage in the material, $D(E)$ is the damage function defined above, and $\phi(E)$ is the fluence of neutrons. The $E_{dis}$ for the various neutron irradiations is found in table \ref{table:neutron_irradiation}.

\subsection{DD Generator Exposures}
\label{appendix:dd}

We used a Thermo Electron Corporation MP320 DD neutron generator to expose our detector to DD neutrons. Nominally, it emits quasi-monoenergetic 2.45 MeV neutrons at a rate of $10^6$ neutrons/second, however, a number of factors decrease the true neutron rate absorbed in the device from this nominal rate. First, the $A =$ 1 cm$^2$ device covers only a small portion of the 4$\pi$ area into which the generator emits neutrons. As the device is approximately $r=$ 6 cm away from the DD interaction point during irradiation, we assume that the fraction of neutrons emitted from the DD interaction point which pass through the detector is $f_\Omega = A/(4 \pi r^2) \approx 0.0022$.

Additionally, the emission from a DD generator is neither isotropic nor independent of the energy of the incoming deuterium beam. \cite{DDGeneratorPaper,DDTheory} The nominal neutron emission rate $R_\mathrm{nom} = 10^6$ neutrons/second is the 4$\pi$ neutron emission rate for the neutron generator running at a voltage of 100 kV, which we actually run at 80 kV to prolong the generator's life, and which interacts with the detector under irradiation approximately 90 degrees from the incoming DD beam. Taking into account both factors using the tables in Ref. \cite{DDTheory}, we calculate that the actual neutron rate at our location and tube voltage is reduced by a factor of
\begin{eqnarray}
    f_{\mathrm{V}\theta} = \frac{4 \pi \frac{\partial \sigma}{\partial \Omega}|_{90^\circ, \mathrm{\ } 80\mathrm{\ keV}}}{\sigma_{tot,\mathrm{\ } 100 \mathrm{\ keV}}} = \frac{4 \pi \times 0.731 \mathrm{\ barn}}{15.2 \mathrm{\ barn}} = 0.604
\end{eqnarray}

Using Geant4, we simulate the incidence of $10^6$ DD generator neutrons on a stack of 50 detectors, and find that $f_\sigma = 0.0091$ of them interact with the silicon detectors (assumed to have a natural isotope ratio), depositing an average of $E_\mathrm{dep}=$ 104 keV per interaction. This compares favorably to analytical estimates for $^{28}$Si, which assume a cross section 2.41 barns. From this, we calculate a rate of 0.0120 Hz, depositing 166 keV per neutron assuming the neutron scatters isotopically and elastically, in reasonable agreement with the simulated data. In the following calculations, we use the simulation data, as it more accurately captures the details of the neutron-silicon interaction. At this energy, we calculate a Lindhard non-ionizing yield of $f_{NI}(104 \mathrm{\ keV}) = 0.502$.

Combining these factors using our Geant4+Lindhard model yields
\begin{eqnarray}
    P_{DD} = R_\mathrm{nom} f_\Omega f_{\mathrm{V}\theta} f_{NI}(104{\mathrm{ keV}}) E_D = 2.27 \mathrm{\ GeV/hr}
\end{eqnarray}
where $P_{DD}$ is the non-ionizing (i.e. potentially damage inducing) power deposited by neutrons in the detector during DD irradiation.

We irradiated one detector in set A with DD neutrons for 140 hours, and one detector in set B for 114.5 hours, yielding a total neutron energy deposition of 318 GeV and 256 GeV for the detectors in set A and B, respectively, using the Geant4 + Lindhard method.

\subsection{AmBe Source Exposure}
\label{appendix:ambe}

In addition to neutrons from the DD generator, we exposed the irradiated detector A to neutrons from an Americium Beryllium (AmBe) alpha-neutron source. While the gamma activity of AmBe sources is relatively easy to measure, the neutron activity of the source is less straightforward to reconstruct. As the AmBe source itself is made of a mixture of Americium and Beryllium powders, the details of e.g. the powder size and degree of mixing affect the neutron yield. To estimate the fast neutron activity of the source, we both extrapolate the neutron activity from the known gamma activity and attempt to measure the neutron activity directly.

Ref. \cite{AmBeStudy} measured the neutron and gamma activities of two AmBe sources, finding a neutron yield of
\begin{eqnarray}
    \phi_{\gamma n} = \frac{\Gamma_{n}}{\Gamma_\gamma} = (6.1 \pm 0.6) \times 10^{-5}
\end{eqnarray}
in units of neutrons per gamma emitted from the source. Our nominal AmBe gamma activity was 51.77~mCi or 1.92~GBq, yielding a nominal neutron activity of $\Gamma_{AmBe}\approx(1.2\pm 0.1)~\times~10^{5}$ neutrons per second (emitted isotropically).

We also measured our source activity directly with a EJ309 organic liquid scintillator detector, using pulse shape discrimination to separate neutrons from gammas interacting with the detector. Using a Geant4 simulation to account for detection efficiencies, we reconstruct a source rate of $\Gamma_{AmBe}\approx (2\pm1)\times10^5$ neutrons per second, in agreement with the estimated neutron activity assumed above. For the following calculations, we take the estimated neutron rate of $\Gamma_{AmBe}\approx(1.2\pm 0.1) \times 10^{5}$ neutrons per second as it gives a more precise nominal neutron rate.

Similarly to the calculation for the DD generator, we need to calculate the fraction of neutrons emitted from the source that pass through the detector. However, in contrast to the pointlike source of the DD generator, the AmBe source is a cylinder with a diameter of 1 cm and a length of 2 cm. During irradiation, the distance between the bottom face of the AmBe source and top of the detector was approximately 0.5~cm. To estimate the number of neutrons emitted from the source which are incident on the detector, we perform a Monte Carlo simulation in which neutrons are emitted from a random position within the source in a random direction, and we count the number of neutrons which pass through the detector under irradiation. We find that fraction of neutrons from the extended AmBe source which pass through the detector is $f_\Omega = 0.0426$, in reasonable agreement with the fraction calculated assuming a pointlike source located at the center of the AmBe cylinder.

To calculate the fraction of neutrons absorbed in the detector and the energy deposited in its nuclear system, we perform a Geant4 simulation using the AmBe spectrum from Ref. \cite{AmBeStudy}. We find that $f_\sigma = 0.00902$ of incident neutrons interact, depositing an average of $E_\mathrm{dep}=$249 keV per interaction. Interpolating the Lindhard model to our average AmBe interaction energy gives $f_{NI}(249 \mathrm{\ keV}) = 0.330$.

Combing these factors we have
\begin{eqnarray}
    P_{AmBe} = \Gamma_{AmBe} f_\Omega E_{D,AmBe} f_{NI}(249 \mathrm{\ keV})\\
    = 13 \pm 1 \mathrm{\ GeV/hr}
\end{eqnarray}

\subsection{Cosmic Ray Neutrons}
\label{appendix:crs}

In this paper, we assume that the majority of the ``natural'' lattice damage in silicon is caused by cosmic ray neutrons, which we assume should dominate over radiogenic neutrons in most environments at the surface. Cosmic rays also contain heavier nuclei (alphas, Li nuclei, etc.) which we do not consider here, and suggest as a topic for future study. Very briefly, we expect heavier nuclei to produce more damage per particle than neutrons (Non-Ionizing Energy Loss, NIEL, for neutrons at 10 MeV is around 1 keV/cm, while for alphas at 10 MeV/nucleon is around 200 keV/cm\cite{NIELCalculator}), but expect that neutrons should still dominate energy deposition in our silicon detectors given their much larger rate in cosmic rays (at sea level, EXPACS\cite{EXPACS1, EXPACS2, EXPACS3, EXPACS4} predicts around 4 orders of magnitude more neutrons than alphas at 10 MeV/nucleon). Ions may produce different damage clusters than neutrons, potentially creating different rate or shape LEE spectra for equivalent amounts of energy deposited in the nuclear system of the crystal.

To make a concrete prediction of the impact of cosmic rays on our devices, we calculate the spectra of cosmic ray neutrons using EXPACS\cite{EXPACS1, EXPACS2, EXPACS3, EXPACS4} at our laboratory at Berkeley (latitude: 38 degrees, longitude: -122 degrees, altitude: 100 meters). To translate this spectrum into energy deposited in our detector, we simulate the mean interaction frequency and energy deposited in the nuclear system of our device at 24 different energies, between 10 keV and 10 GeV (the range where most cosmic ray neutron energy is deposited), and linearly interpolate these to model the fractions of incident neutrons which interact $f_\sigma$ and the energy deposited in the detector for a neutron at a given energy $E_{dep}(E_n)$. We include the modeled Lindhard non-ionizing yield to find the total energy deposited per neutron at a given energy incident on the detector $E_D(E_n) = E_{dep}(E_n) f_{NI}(E_{dep}(E_n)) f_\sigma(E_n)$.

\begin{figure}
\includegraphics[width=1\columnwidth]{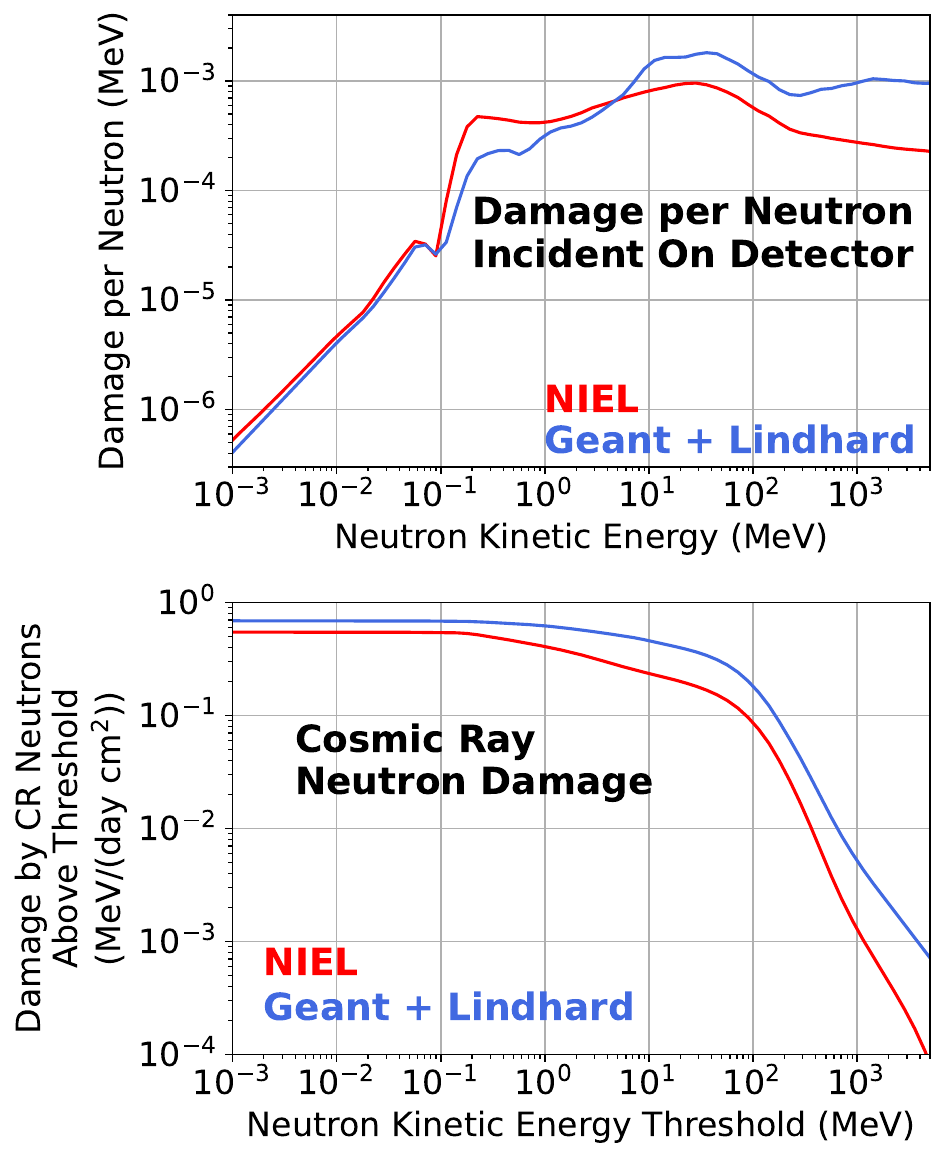}
\caption{\label{fig:cr_sim} (Top) Damage, i.e. energy deposited in the nuclear channel, per neutron incident on the detector at a given energy modeled with a NIEL model\cite{NIELSi} and a model based on Geant4 simulations and with a Lindhard correction\cite{Lindhard1961} (see text). (Bottom) Integrated damage from cosmic ray neutrons above a threshold (x-axis), using EXPACS\cite{EXPACS1, EXPACS2, EXPACS3, EXPACS4} to model cosmic ray neutron fluxes at the surfaces.}
\end{figure}

To gain intuition about which neutron energies contribute most to the total energy deposited in the silicon detector, we integrate the neutron flux $d \phi(E)/dE$ above a threshold and convolve it with the modeled fraction of energy deposited in the detector
\begin{eqnarray}
    E_{dep}(E_{thresh}) = \int_{E_{thresh}}^\infty \frac{d \phi}{dE_n}(E_n) f_\sigma(E_n) E_{dep}(E_n) dE_n
\end{eqnarray}
which we plot in Fig. \ref{fig:cr_sim}.

We calculate that on the ground, our cosmic ray neutrons deposit approximately 0.69 MeV per day in our detector (using the Geant4 + Lindhard model). To estimate the energy deposited while the detectors are in transit via plane, we consider the cosmic ray neutron spectrum at 35,000 feet above sea level and find they deposit roughly 4.6 MeV per hour. The full location history of the detector silicon can be found in table~\ref{tab:cr_history}.

\begin{table*}[h]
\caption{\label{tab:cr_history} History of detector substrates, which is used to construct the cosmic ray exposure model. While the fabrication date is known to be in 2023, the exact fabrication date is not known. Here, we assume the detector was manufactured on June 1, 2023. The total time on ground or on planes is for time between boule manufacturing and the start of the set B detectors run. Note that the detectors in Set B were run chronologically before the detectors in Set A.}
\begin{tabular}{|c|c|c|c|}
\hline
Operation & Location & Ground/Plane & Duration \\
\hline
Crystal fabrication & Fujian, China & Ground & 300 d \\
\hline
Boule shipping & Fujian, China to New Jersey, USA & Plane & 19.1 hr \\
\hline
Slicing, polishing, storage & New Jersey, USA & Ground & 847 d \\
\hline
Wafer shipping & New Jersey, USA to LBNL, USA & Plane & 6.75 hr \\
\hline
Wafer storage & LBNL, USA & Ground & 13 d \\
\hline
Wafer shipping & LBNL, USA to TAMU, USA & Plane & 4 hr \\
\hline
Device microfabrication & TAMU, USA & Ground & 42 d \\
\hline
Shipping to dicing & TAMU, USA to San Jose, USA & Plane & 4 hr \\
\hline
Dicing & San Jose, USA & Ground & 9 d \\
\hline
Shipping to inspection & San Jose, USA to TAMU, USA & Plane & 4 hr \\
\hline
Inspection & TAMU, USA & Ground & 6 d \\
\hline
Shipping to Berkeley & TAMU, USA to LBNL, USA & Plane & 4 hr \\
\hline
Storage Until Running Set B & LBNL, USA & Ground & 151 d \\
\hline
\hline
Total time on ground &  & Ground & 1386 d \\
\hline
Total time on planes &  & Plane & 41.85 hr \\
\hline
\hline
Storage Set B run to Set A run 1 & LBNL, USA & Ground & 28 d \\
\hline
\end{tabular}
\end{table*}

\begin{table}[h]
\caption{\label{tab:irradiation_time} Summary of the length of each neutron irradiation and the length of time between end of irradiation and the start of the run (approximately when the detector reached 4 K, ``Time Until Cold''). During irradiation and before being mounted in the fridge, the detectors were kept at room temperature. The cooldown to mK temperatures took approximately 36-48 hours.}
\begin{tabular}{|c|c|c|c|c|c|}
\hline
\textrm{Run} & Irradiation Type & \textrm{Irradiation Time} & \textrm{Time Until Cold}\\
\hline
Set A Run 1 & DD & 140 hr & 17.0 d  \\
Set A Run 1 & AmBe & 24 hr & 9.75 d  \\
\hline
Set B & DD & 114.5 hr & 3.5 d \\
\hline
\end{tabular}
\end{table}

\section{Vibration Induced Backgrounds and Cuts}
\label{appendix:vibration_backgrounds}

\begin{figure}
\includegraphics[width=1\columnwidth]{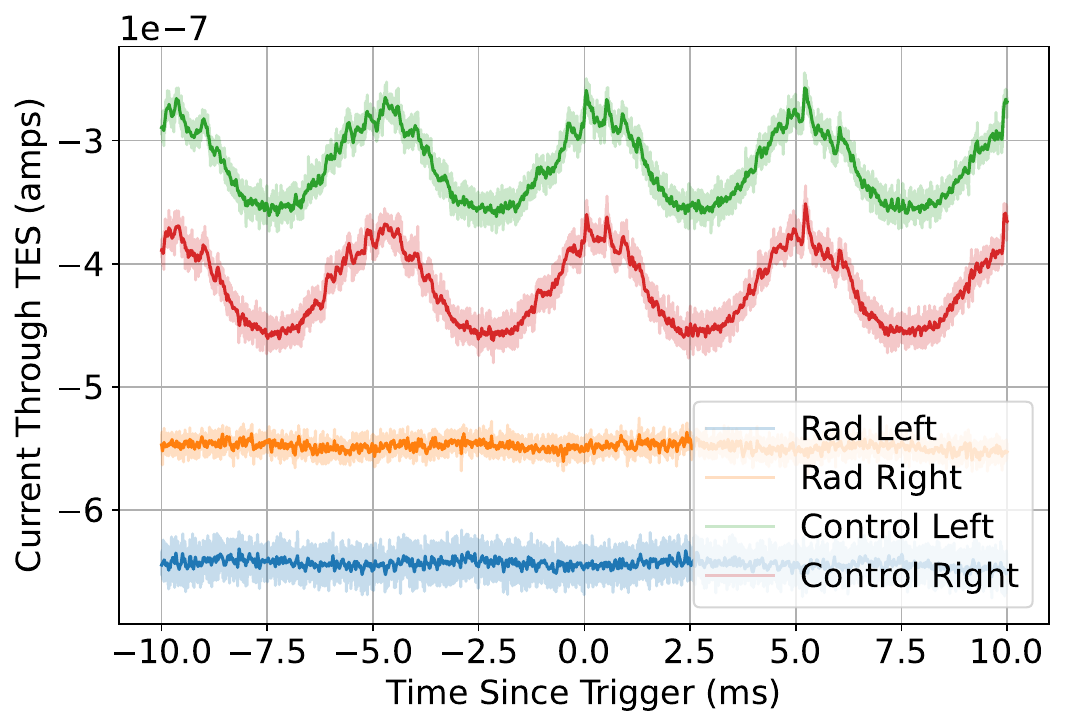}
\caption{\label{fig:vibrations_events} Example of a vibration induced low energy event in the control device in the second run of the set A detectors. These vibration events occur around the maximum of the quasisinusoidal vibration induced low frequency noise (where mechanical loss associated with the detector's vibration produces large amounts of power loading). The rate of events around the minima of the vibration noise is significantly smaller. Note that in addition to the triggered event at 0 ms, there are additional vibration induced events, e.g. between 0 and 1 ms and around 5.2 ms. These vibration induced events are correlated between two channels of one detector, but are uncorrelated with events in the other detector.}
\end{figure}

As briefly noted in the main text, we have observed low energy phonon events created in phase with the vibration of the wire bonds used to suspend our detectors (see Fig. \ref{fig:vibrations_events}). Briefly, we believe the release of mechanical stress energy in the bond foot of the structure is responsible for this effect. While this effect will be described in detail in future work, here, we are primarily concerned with cutting these events that are unrelated to the LEE effect we are attempting to study. 

\begin{figure}
\includegraphics[width=1\columnwidth]{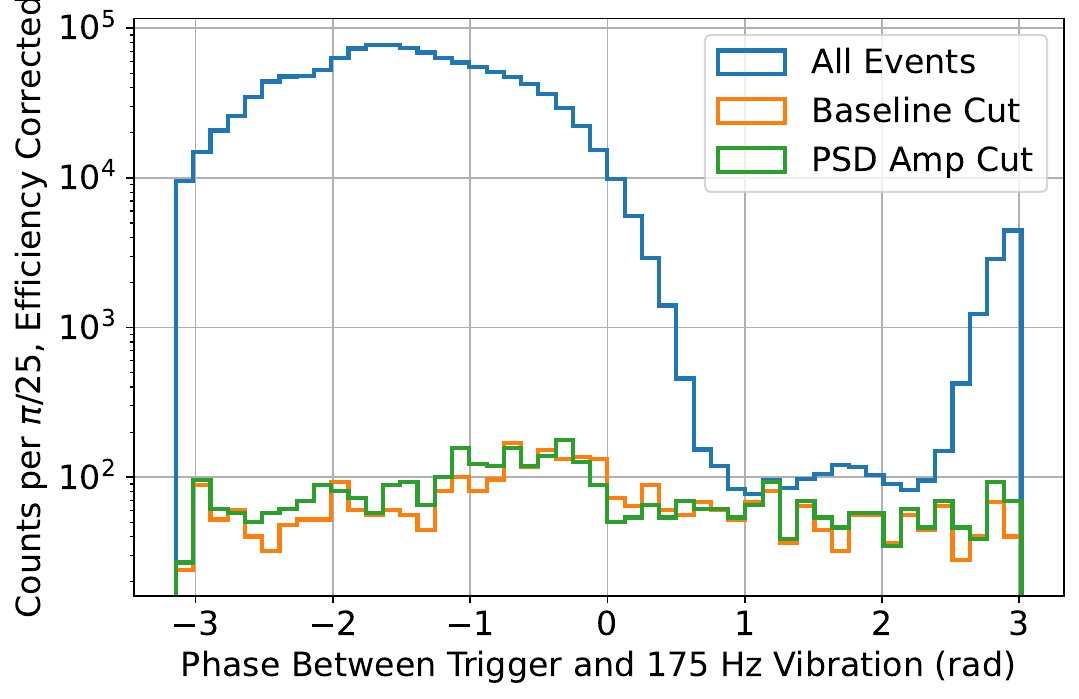}
\caption{\label{fig:vibrations_phase} Example of a histogram of phase (timing of event trigger relative to the 175 Hz mechanically induced vibrations) in the control detector in the second run of the set A of detectors. To ensure that we are not purely measuring a broad decrease in the rate of events, but instead preferentially cutting vibration induced events in the phase peak, we correct for the efficiency of the cut we apply by dividing by the passage fraction of the cut on events chosen at random times throughout the dataset. The ``baseline'' and ``PSD amp'' cuts are described in the text. }
\end{figure}

In particular, we observe that in many datasets, low energy events tend to occur around the maximum amplitude of the main quasisinusoidal noise caused by mechanical vibration of the detector. When plotting a histogram (see Fig. \ref{fig:vibrations_phase}) of the observed phase difference between this main vibration peak and the event trigger (i.e. the time difference over the vibration peak period), we observe a very large excess at certain phases. Using this feature as a herald of vibration induced events, we construct cuts that remove periods of time with high levels of vibration. We employ two approaches as described below.

First, periods with high vibrations increase the average DC power loading on the detector, and therefore increase the average pre-pulse ``baseline.'' We design cuts which remove events where this baseline exceeds a set value, and progressively reduce this set value until the vibration induced peak is removed from the phase histogram.

Similarly, when Fourier transforming traces from periods of high vibration, there will be a peak at the frequency of the vibration noise, with the height proportional to the amplitude of the vibration induced noise. We cut on the amplitude of this peak (the ``PSD amp'') to remove time periods when there is a large amount of vibration induced noise and therefore a high rate of vibration induced low energy events. Similar to the baseline cut, we cut events for which the amplitude of a Fourier transform of a pre-pulse region at the main vibration frequency exceeds a threshold, and set this threshold such that we both remove the main vibration induced peak seen in the phase histogram while passing as many events as possible. 

We find that these two independent cut methods remove very similar amounts of events, producing largely similar phase and energy spectra. We use the baseline cut as our method for removing vibration induced events in all our datasets, as this cut appears to be slightly more efficient at removing vibration induced events while passing non-vibration events.

\section{Coincident Backgrounds and Cuts}
\label{appendix:coincident_backgrounds}

\begin{figure}
\includegraphics[width=1\columnwidth]{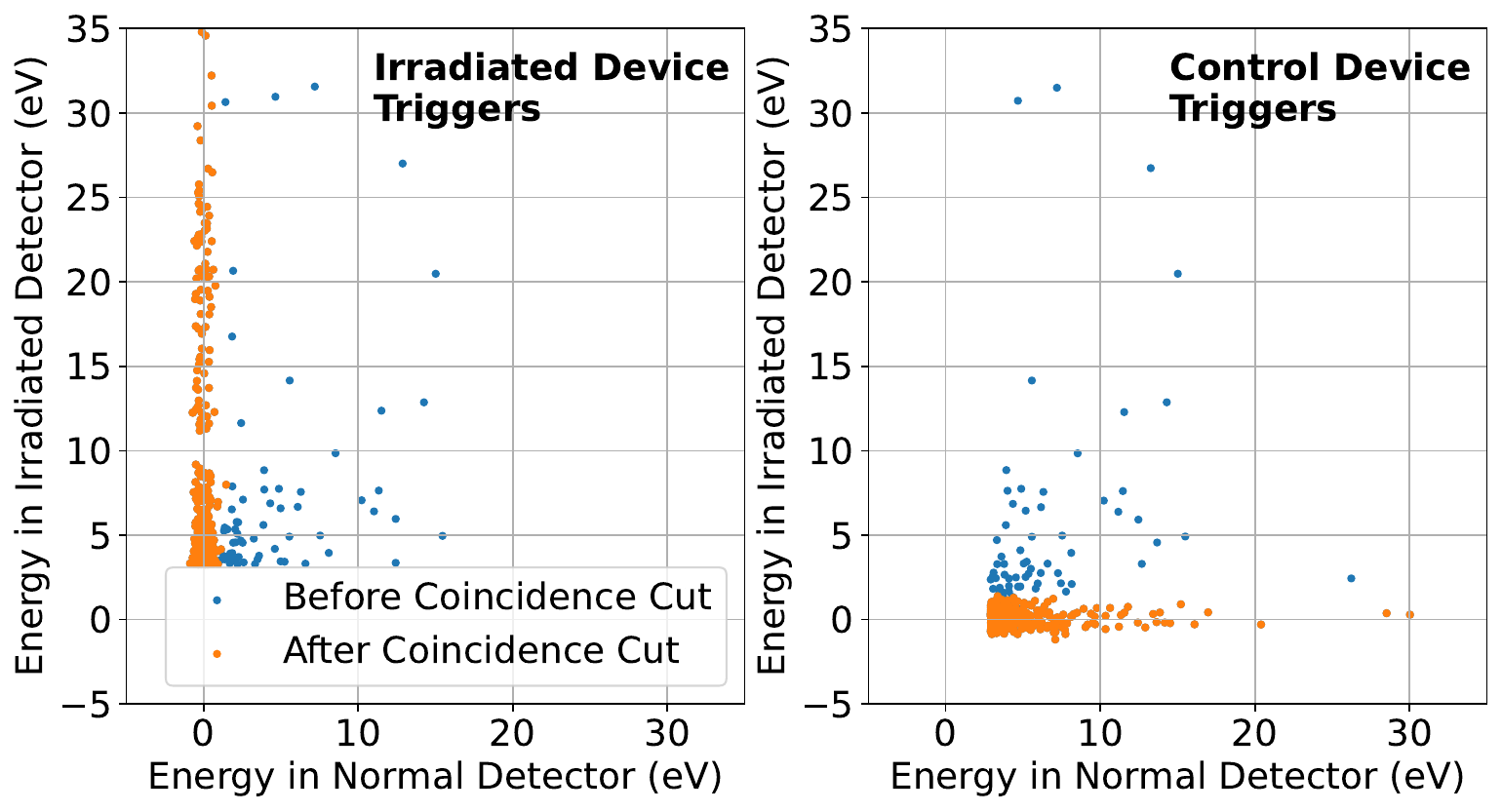}
\caption{\label{fig:coincidence} Energies of triggered events in the second dataset of the first run measuring detectors in Set A, assuming a phonon-like pulse shape. (Left) shows events triggered in the irradiated detector, while (Right) shows events triggered in the control detector. Blue shows events removed via the coincidence cut, while orange shows events passing the coincidence cut (see text).}
\end{figure}

In our data, we observe events which are coincident between both detectors (see Fig. \ref{fig:coincidence}), which seem to come in at least two varieties.

\begin{figure}
\includegraphics[width=1\columnwidth]{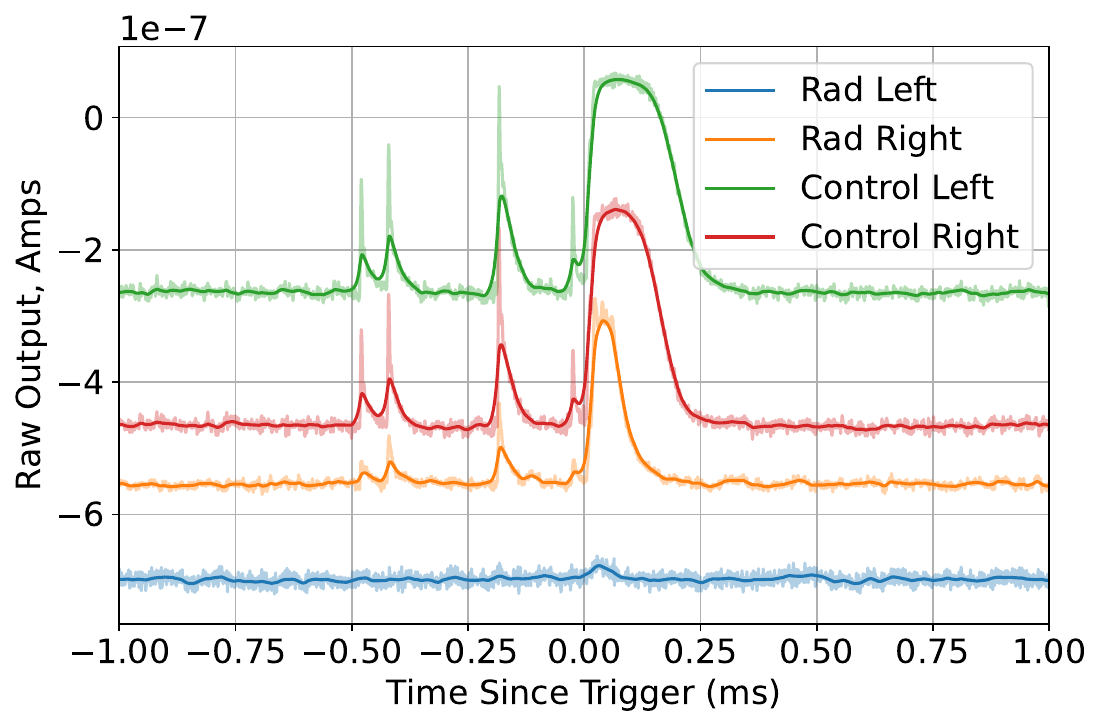}
\caption{\label{fig:coincidence_emi} ``Coincident'' event with a non-phonon pulse shape. Light lines are the raw waveforms, and dark lines are filtered with a 20 kHz low pass filter for clarity. Green and red (blue and orange) are the left and right channels of the control (irradiated) detector. We hypothesize that these events are caused by EMI (see text).}
\end{figure}

First, we seem to observe events with obviously non-phonon pulse shapes, coupling at different strengths to different channels (see Fig. \ref{fig:coincidence_emi}). We interpret these as events created by electromagnetic interference (EMI) being injected down our TES bias lines, heating our TESs and creating a non-phonon pulse shape.

\begin{figure}
\includegraphics[width=1\columnwidth]{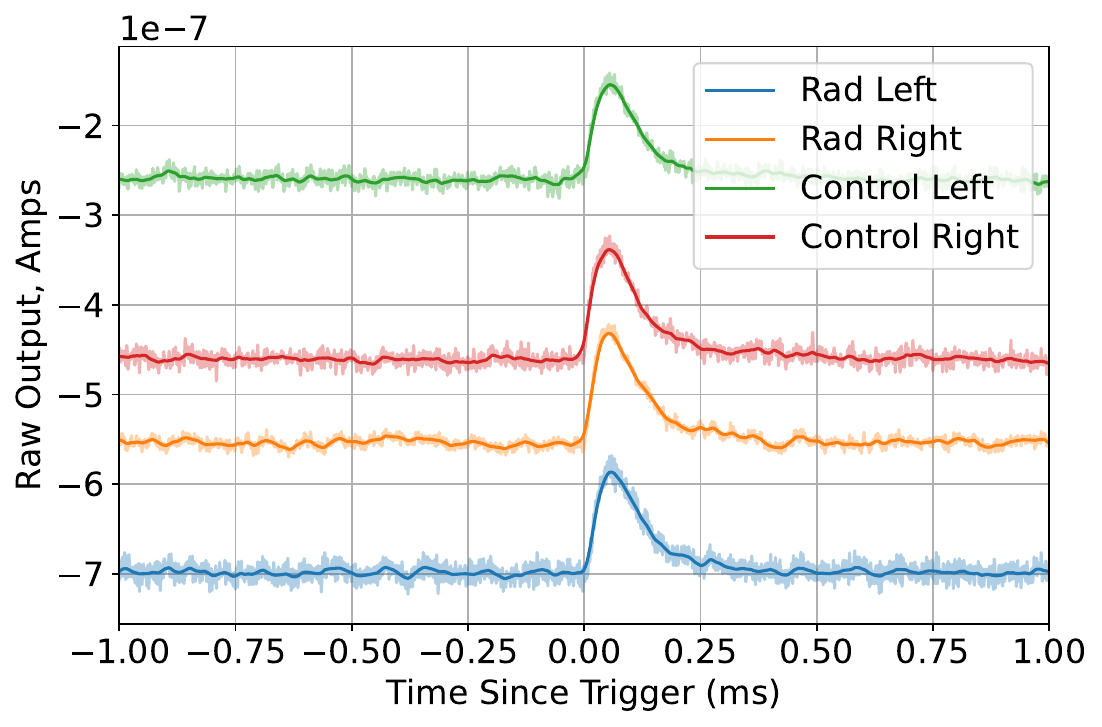}
\caption{\label{fig:coincidence_pcb} ``Coincident'' event with a phonon-like pulse shape. Light lines are the raw waveforms, and dark lines are filtered with a 20 kHz low pass filter for clarity. Green and red (blue and orange) are the left and right channels of the control (irradiated) detector. We hypothesize that these events are caused by our PCB scintillating (see text).}
\end{figure}

Second, we observe events with phonon-like pulse shapes that couple approximately equally to both channels of a single detector, but with varying strengths between the two detectors. Because both detectors are operated within a single optical cavity adjacent to a printed circuit board (PCB), scintillation from the PCB is expected to produce optical phonon bursts that generate coincident events in different detectors, as discussed in Ref. \cite{CDMSHVeVR3}. An example waveform is seen in Fig.~\ref{fig:coincidence_pcb}.

Since neither event type is related to the phonon-only LEE process under study, we reject these events using a coincident cut. This cut accepts events only when the no-delay amplitude from the two-channel, single-amplitude optimal filter fit in the other detector is less than 4$\sigma$ from zero, i.e., consistent with no signal.

\section{Data Quality Cuts}
\label{appendix:quality_cuts}

In addition to the coincident event and vibration induced event cuts, we employ two additional cuts to ensure that the events accepted into the final analysis are consistent with phonon-coupled LEE events.

\begin{figure}
\includegraphics[width=1\columnwidth]{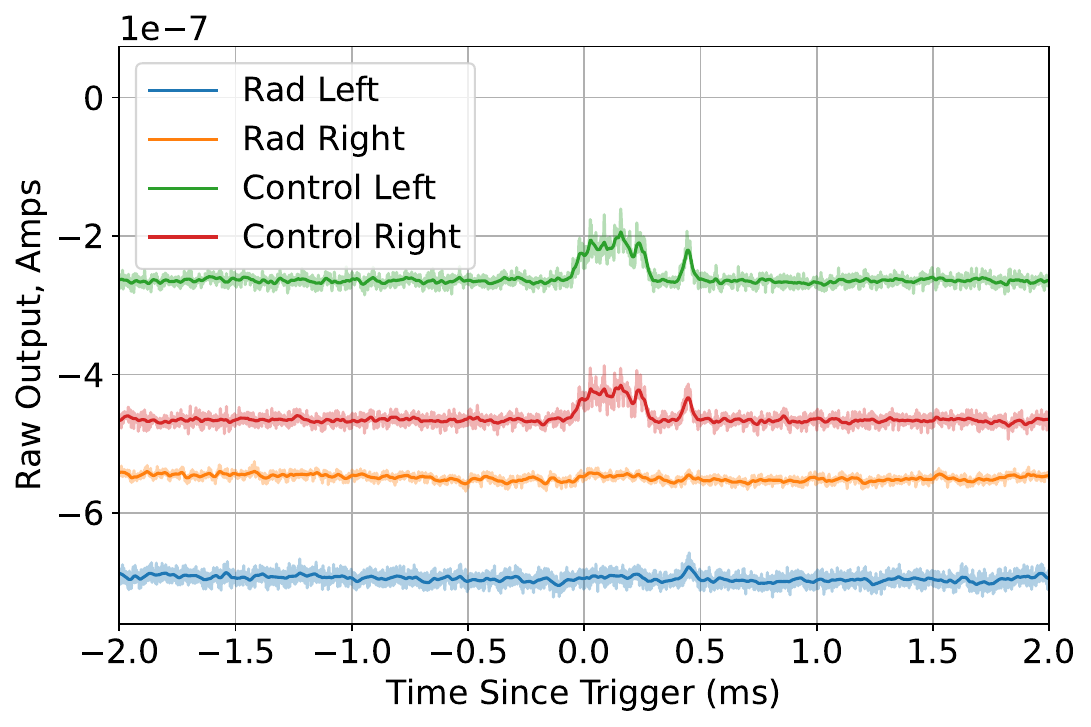}
\caption{\label{fig:chi2_example} Example of an event failing the low frequency $\chi^2$ cut, but passing all other cuts, observed in the control device in the second run of the set A detectors. The event has a clearly non-phonon induced event shape, leading us to hypothesize that EMI may be responsible for these events.}
\end{figure}

First, we apply a low frequency $\chi^2$ cut, which compares the observed shape of the events to a phonon event template, considering only frequencies below 50 kHz, where signal dominates. This cut removes events for which this low frequency $\chi^2$ exceeds an amplitude dependent threshold, i.e. which are not shaped like phonon events. While other cuts successfully remove the vast majority of events which would fail this low frequency $\chi^2$ cut, the cut does remove a small number of unusually shaped pulses, possibly induced by EMI (see Fig. \ref{fig:chi2_example}).

Finally, we apply a $\delta \chi^2$ based cut to remove ``single'' events \cite{TwoChannelPaper, CRESSTTwoChannel} (i.e. those occurring within the aluminum fins of our detector \cite{TwoChannelPaper} rather than within the substrate), identically to as described in Refs. \cite{TwoChannelPaper, TwoChannelLimits, Run57Paper}. In short, we find the difference between the $\chi^2$ from fitting two-channel, single-amplitude templates to both channels of a single detector using either a phonon template or a ``singles'' template (in which a single faster pulse occurs in one channel, with essentially no response in the other). Through making these comparisons, we can determine if an event is more ``phonon-like'' or ``singles-like,'' considering both the pulse shape and relative response in each channel.

\begin{figure}
\includegraphics[width=1\columnwidth]{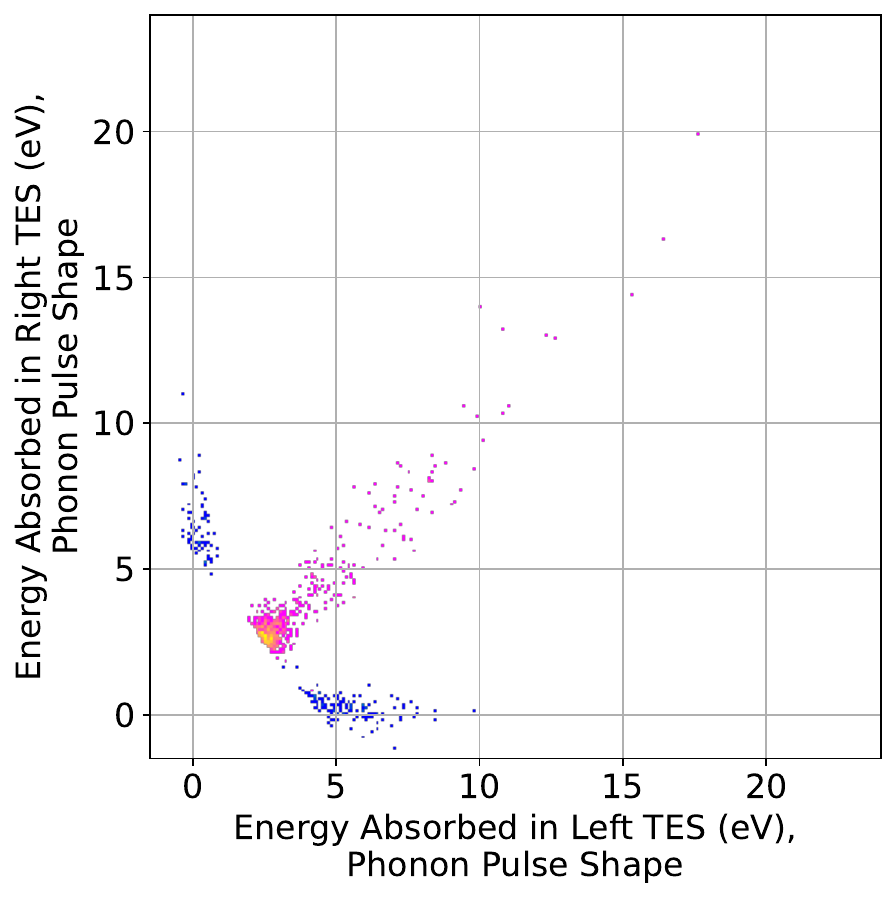}
\caption{\label{fig:singles_shared} Scatter plot of triggered events in the control detector in the second run of the set A detectors, showing energy absorbed in the left vs. right TES, assuming a phonon pulse shape. Events passing the $\delta\chi^2$ based single/shared cut (i.e. shared/phonon events) are shown in pink, while events failing the cut (i.e. singles) are shown in blue. Note that our trigger threshold in this work is somewhat higher than in previous publications. }
\end{figure}

\section{Cut Passage Efficiency}
\label{appendix:cut_efficiency}

The acceptance of the above quality cuts is measured at a range of energies: 0 eV, 15 eV, 20 eV, and 25 eV as these sampled the region of interest in this work, and it allows us to confirm the acceptance has no energy dependence. The acceptance at 0 eV is measured via a population of randomly sampled events when the cuts are applied. These events, being random in nature, will consist of a largely quiescent baseline and its fluctuations and are unlikely to have real phonon events. However, given that the rate of events is non-zero, there is still some probability that a random selection has a pulse in it, but given the rate is low, the large majority of these events will have a maximum reconstructed event amplitude Gaussian distributed around zero with the $\sigma$ of the distribution given by the detector resolution. To measure the cut acceptance, we take the ratio of events cut from these samples to the total number of events. The acceptance at energies above 0 eV is done via "salting". Here, we inject fake signals into the data stream at a known rate and energy and measure the passage fraction just like the randomly selected events. Though these events have non-zero energy, they are randomly distributed in time just as the randomly selected events. More information on the salting algorithm used can be found in~\cite{TwoChannelLimits}. The passage efficiency of the various cuts described in \ref{appendix:coincident_backgrounds},\ref{appendix:vibration_backgrounds}, and \ref{appendix:quality_cuts} can be found below in tables \ref{tab:eff_setA_run1},\ref{tab:eff_setA_run2}, and \ref{tab:eff_setB}.

\begin{table*}[ht]
\caption{Cut efficiencies for Set A Run 1.}
\centering
\resizebox{\textwidth}{!}{%
\begin{tabular}{|l|c|c|c|c|c|c|c|c|}
\hline
 & \multicolumn{8}{c|}{\textbf{Cut efficiencies -- Set A Run 1}} \\
\hline
 & \multicolumn{4}{c|}{Irradiated} & \multicolumn{4}{c|}{Control} \\
\hline
 & 0 eV & 15 eV & 20 eV & 25 eV 
 & 0 eV & 15 eV & 20 eV & 25 eV \\
\hline

Low Frequency $\chi^2$ Cut 
& $75.4\% \pm 0.04\%$ & $75.2\% \pm 0.3\%$ & $75.2\% \pm 0.3\%$ & $75.2\% \pm 0.3\%$
& $88.8\% \pm 0.03\%$ & $88.2\% \pm 0.4\%$ & $88.6\% \pm 0.4\%$ & $88.6\% \pm 0.4\%$ \\
\hline

Quiescent Baseline Cut 
& $82.5\% \pm 0.04\%$ & $82.2\% \pm 0.4\%$ & $82.1\% \pm 0.4\%$ & $82.3\% \pm 0.4\%$
& $91.9\% \pm 0.03\%$ & $91.4\% \pm 0.4\%$ & $91.6\% \pm 0.4\%$ & $91.6\% \pm 0.4\%$ \\
\hline

Coincidence Cut 
& $99.6\% \pm 0.01\%$ & $99.0\% \pm 0.4\%$ & $99.1\% \pm 0.4\%$ & $99.1\% \pm 0.4\%$
& $99.4\% \pm 0.01\%$ & $98.7\% \pm 0.4\%$ & $98.8\% \pm 0.4\%$ & $98.7\% \pm 0.4\%$ \\
\hline
\end{tabular}%
}
\label{tab:eff_setA_run1}
\end{table*}

\begin{table*}[ht]
\caption{Cut efficiencies for Set A Run 2.}
\centering
\resizebox{\textwidth}{!}{%
\begin{tabular}{|l|c|c|c|c|c|c|c|c|}
\hline
 & \multicolumn{8}{c|}{\textbf{Cut efficiencies -- Set A Run 2}} \\
\hline
 & \multicolumn{4}{c|}{Irradiated} & \multicolumn{4}{c|}{Control} \\
\hline
 & 0 eV & 15 eV & 20 eV & 25 eV 
 & 0 eV & 15 eV & 20 eV & 25 eV \\
\hline

Low Frequency $\chi^2$ Cut 
& $93.3\% \pm 0.05\%$ & $93.1\% \pm 0.4\%$ & $93.2\% \pm 0.4\%$ & $92.9\% \pm 0.4\%$
& $47.5\% \pm 0.01\%$ & $47.0\% \pm 0.3\%$ & $47.1\% \pm 0.3\%$ & $46.7\% \pm 0.3\%$ \\
\hline

Quiescent Baseline Cut 
& $99.9\% \pm 0.01\%$ & $99.5\% \pm 0.4\%$ & $99.5\% \pm 0.4\%$ & $99.5\% \pm 0.4\%$
& $31.7\% \pm 0.01\%$ & $31.6\% \pm 0.3\%$ & $31.6\% \pm 0.3\%$ & $31.4\% \pm 0.3\%$ \\
\hline

Coincidence Cut 
& $93.1\% \pm 0.05\%$ & $92.9\% \pm 0.4\%$ & $92.7\% \pm 0.4\%$ & $92.9\% \pm 0.4\%$
& $94.9\% \pm 0.01\%$ & $94.1\% \pm 0.7\%$ & $94.7\% \pm 0.5\%$ & $94.7\% \pm 0.4\%$ \\
\hline
\end{tabular}%
}
\label{tab:eff_setA_run2}
\end{table*}

\begin{table*}[ht]
\caption{Cut efficiencies for Set B.}
\centering
\resizebox{\textwidth}{!}{%
\begin{tabular}{|l|c|c|c|c|c|c|c|c|}
\hline
 & \multicolumn{8}{c|}{\textbf{Cut efficiencies -- Set B}} \\
\hline
 & \multicolumn{4}{c|}{Irradiated} & \multicolumn{4}{c|}{Control} \\
\hline
 & 0 eV & 15 eV & 20 eV & 25 eV 
 & 0 eV & 15 eV & 20 eV & 25 eV \\
\hline

Low Frequency $\chi^2$ Cut 
& $97.9\% \pm 0.03\%$ & $97.6\% \pm 0.4\%$ & $97.6\% \pm 0.4\%$ & $97.6\% \pm 0.4\%$
& $63.3\% \pm 0.01\%$ & $63.5\% \pm 0.4\%$ & $62.9\% \pm 0.4\%$ & $63.0\% \pm 0.3\%$ \\
\hline

Quiescent Baseline Cut 
& $99.8\% \pm 0.01\%$ & $99.5\% \pm 0.4\%$ & $99.5\% \pm 0.4\%$ & $99.4\% \pm 0.4\%$
& $50.3\% \pm 0.01\%$ & $50.5\% \pm 0.3\%$ & $49.8\% \pm 0.3\%$ & $49.9\% \pm 0.3\%$ \\
\hline

Coincidence Cut 
& $97.5\% \pm 0.03\%$ & $96.9\% \pm 0.4\%$ & $96.9\% \pm 0.4\%$ & $96.9\% \pm 0.4\%$
& $98.4\% \pm 0.02\%$ & $98.3\% \pm 0.4\%$ & $98.2\% \pm 0.4\%$ & $98.2\% \pm 0.4\%$ \\
\hline
\end{tabular}%
}
\label{tab:eff_setB}
\end{table*}

\section{Complete Spectra}

\begin{figure}[h]
\includegraphics[width=1\columnwidth]{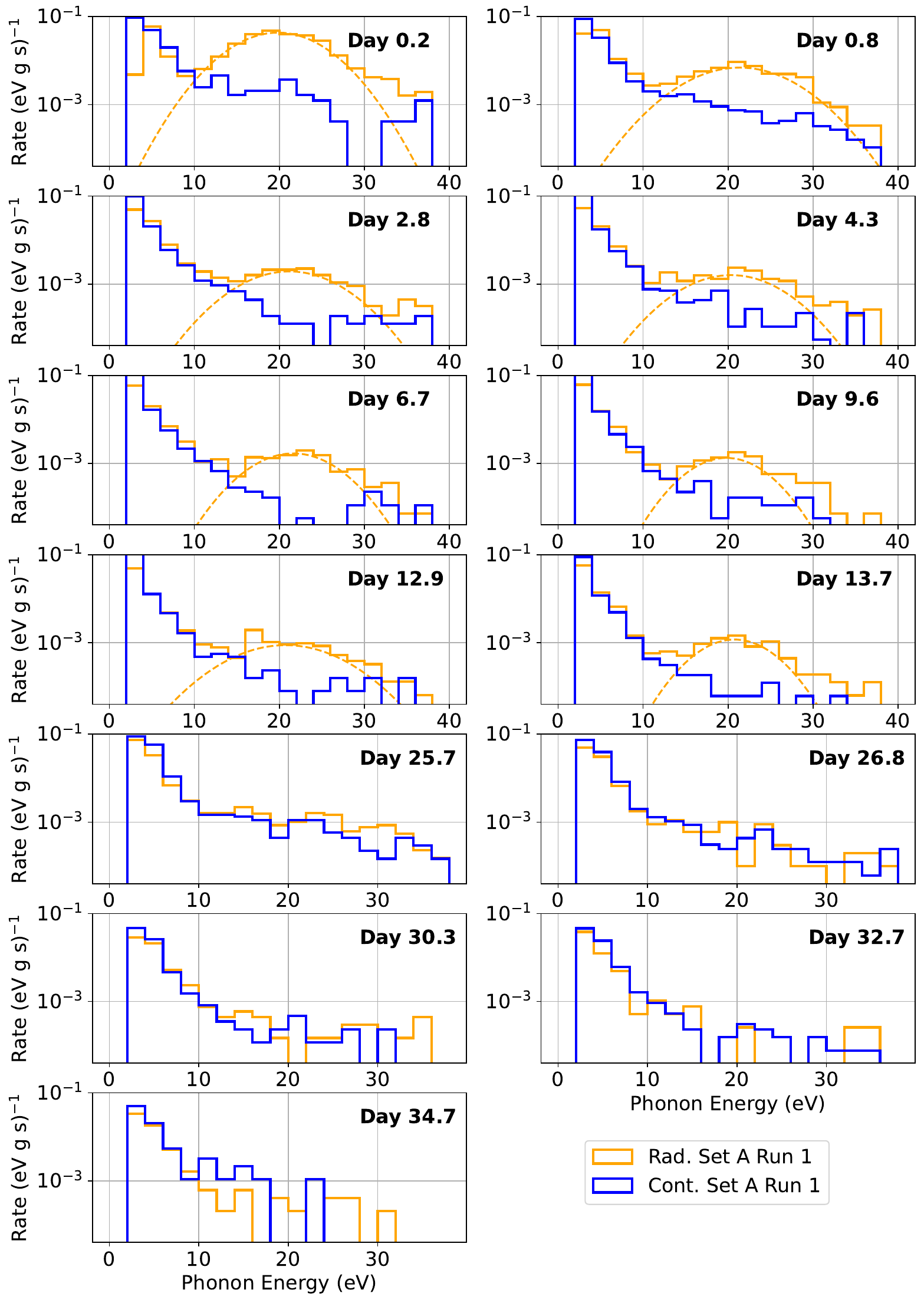}
\caption{\label{fig:all_spectra_r62} Spectra observed in the Set A irradiated and control detectors for each dataset recorded during the first run of the Set A devices. The dashed yellow line shows a fit to the Gaussian peak discussed in the main text, serving to guide the eye. After the 50 K warm up, (day 25.7 and after), this Gaussian feature does not seem to appear in the data, and we do not attempt to fit a Gaussian model.}
\end{figure}

\begin{figure}[h]
\includegraphics[width=1\columnwidth]{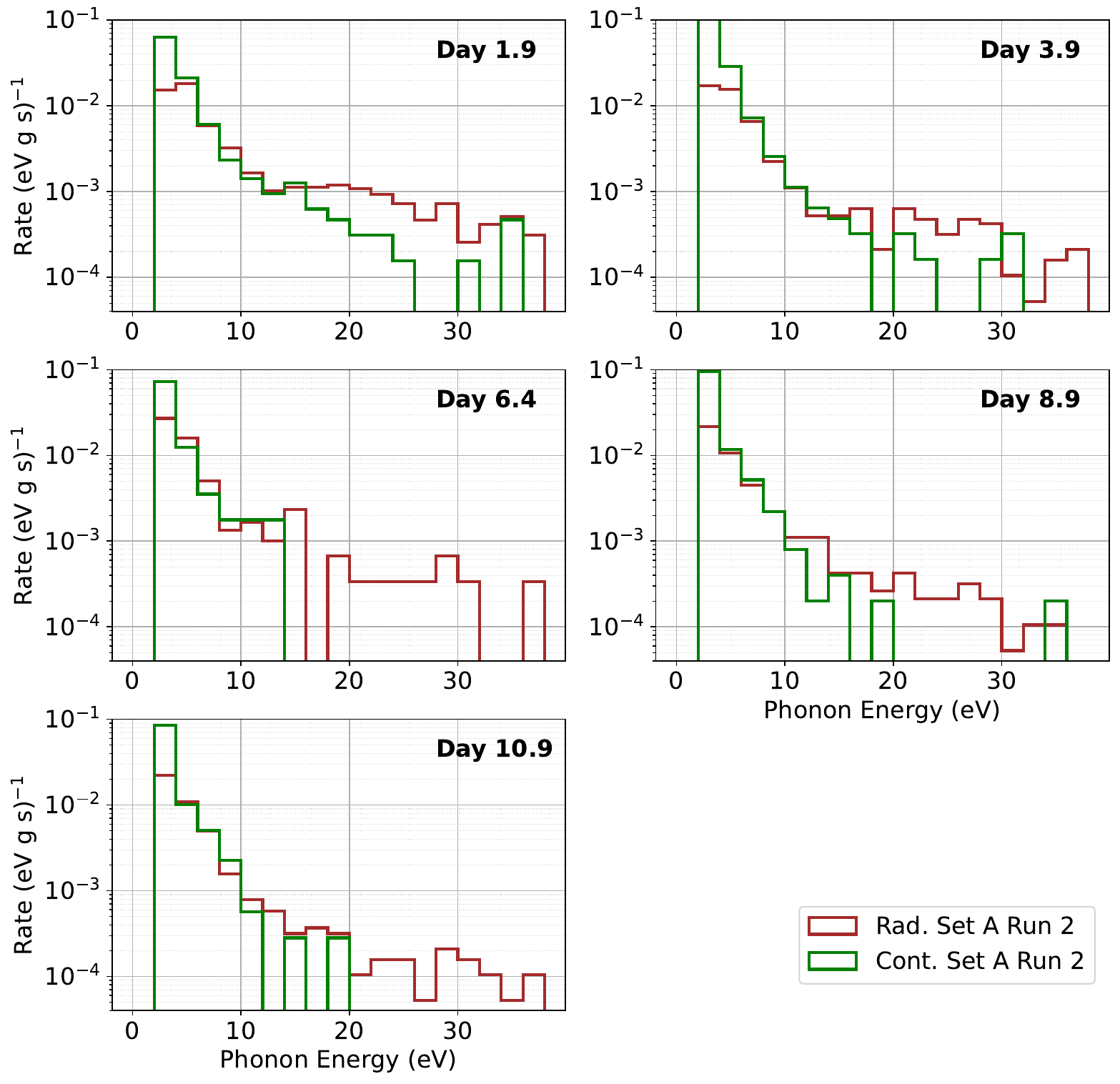}
\caption{\label{fig:all_spectra_r65} Spectra observed in the Set A irradiated and control detectors for each dataset recorded during the second run of the Set A devices.}
\end{figure}

\begin{figure}[h]
\includegraphics[width=1\columnwidth]{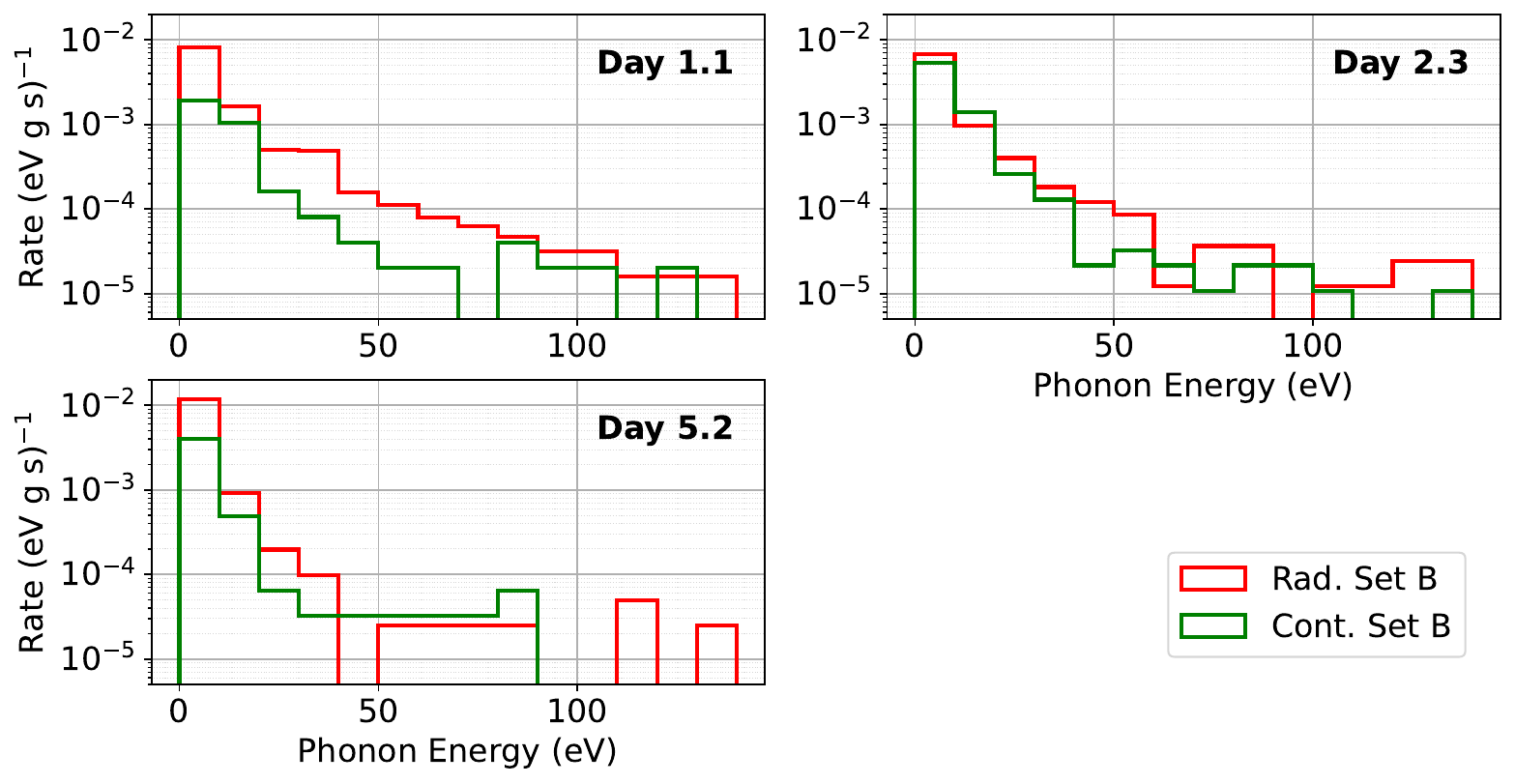}
\caption{\label{fig:all_spectra_r60a} Spectra observed in the Set B irradiated and control detectors for each dataset.}
\end{figure}

\begin{figure}[h]
\includegraphics[width=1\columnwidth]{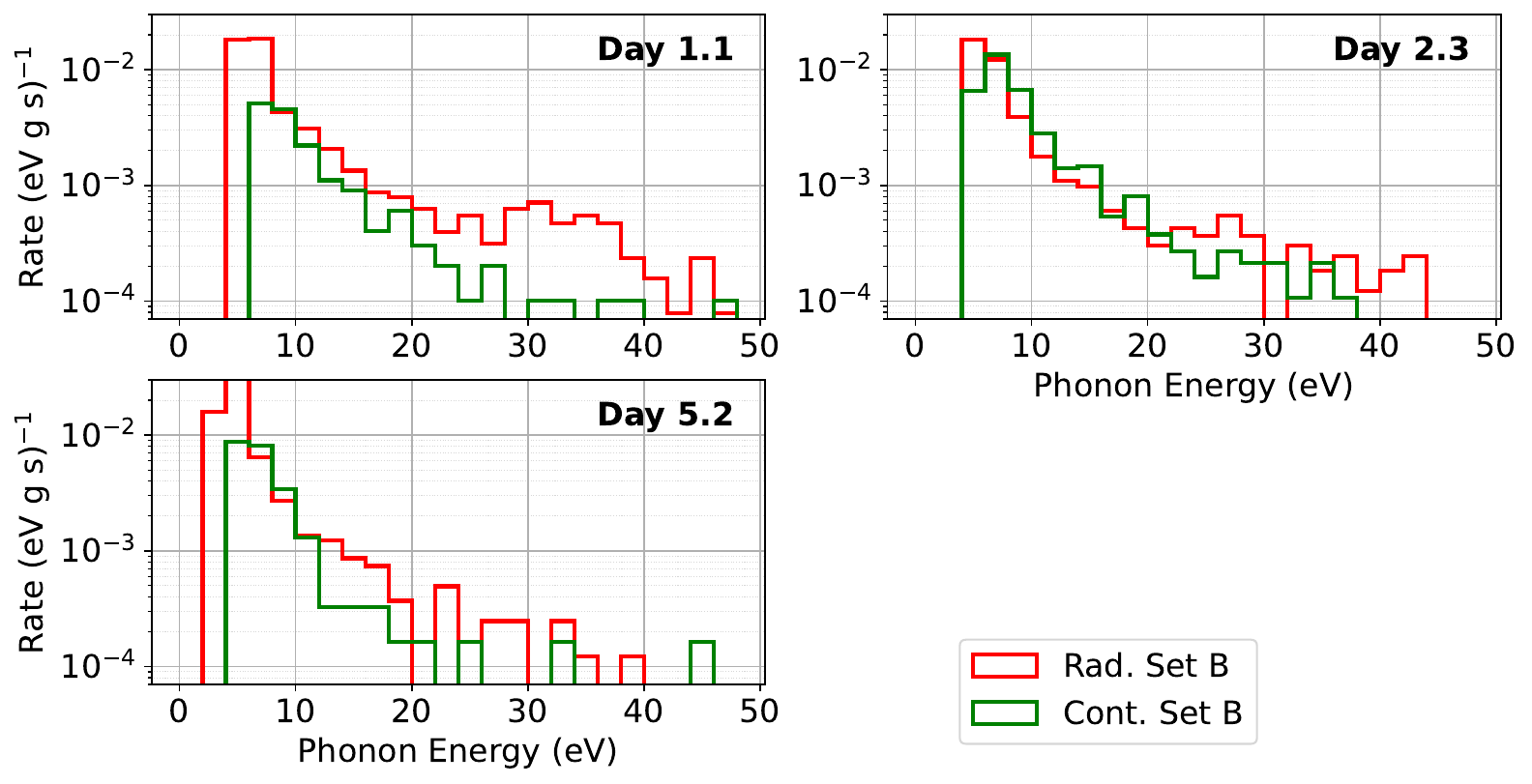}
\caption{\label{fig:all_spectra_r60b} Spectra observed in the Set B irradiated and control detectors for each dataset, showing energies between 0 and 50 eV for comparison to the Set A detector data.}
\end{figure}

In the main text, for the sake of brevity we only show two representative spectra. In Figs. \ref{fig:all_spectra_r62}, \ref{fig:all_spectra_r65}, \ref{fig:all_spectra_r60a}, and \ref{fig:all_spectra_r60b} we show spectra of all the datasets taken as part of this study.

\section{Time Domain Models and Fits}

In the main text (eqns. \ref{eqn:pwr_law}, \ref{eqn:pwr_law_warmup}, and \ref{eqn:pwr_law_fix}), we give three power law models used to fit the low energy phonon event rate as a function of time: a power law with two free parameters, a power law with a modified start time (corresponding to the end of the warm up period), and a power law with a fixed exponent.
\begin{eqnarray}
    R_p(t) = \beta t^{-\alpha} \\
    R_{\mathrm{wu}}(t) = \beta (t - t_{\mathrm{wu}})^{-\alpha} \\
    R_f(t) = \beta t^{-\alpha_{rad}}
\end{eqnarray}

In the first run of the set A detectors, we fit the third through eighth data point in the pre-warm up period with the two-parameter $R_p(t)$ model. The first two data points appear to follow a different trend, which we were unable to satisfactorily fit by introducing a variable start time for the power law (i.e. fitting to $R(t) = \beta (t - t_0)^{-\alpha}$). Certainly, a more complicated model involving e.g. multiple power law terms (e.g. $R(t) = \beta_1 t^{-\alpha_1} + \beta_2 t^{-\alpha_2} $) could fit these early data points, however, adding more free parameters to the model seems to fail to provide much explanatory power. Similarly, for the second run of the set A detectors, we drop the first point from these fits, as it appears to be systematically higher. Although they broadly seem consistent with the same power law constant, the Set B detectors were only measured at three points in time, too few to fit a two parameter power law. The fit parameters for this power law model (\ref{eqn:pwr_law}) are given in table \ref{tab:fits_pwr}.

\begin{table}[h]
\caption{\label{tab:fits_pwr} Power law (eqn. \ref{eqn:pwr_law}) fit parameters for the first and second run of the Set A detectors. Within uncertainties, the power law parameter $\alpha$ is consistent between the irradiated and non-irradiated detectors for both runs, while the prefactor $\beta$ is significantly elevated in the irradiated detector.} 
\begin{tabular}{|c|c|c|c|}
\hline
\textrm{Run} & Detector & Parameter & Fit Value\\
\hline
Set A Run 1 & Irradiated & $\alpha$ & 0.40 $\pm$ 0.06\\
Set A Run 1 & Control & $\alpha$ & 0.4 $\pm$ 0.2 \\
Set A Run 2 & Irradiated & $\alpha$ & 0.7 $\pm$ 0.2 \\
Set A Run 2 & Control & $\alpha$ & 1.1 $\pm$ 0.7 \\
\hline
Set A Run 1 & Irradiated & $\beta$ & 0.038 $\pm$ 0.004 Hz/gram \\
Set A Run 1 & Control & $\beta$ & 0.005 $\pm$ 0.002 Hz/gram \\
Set A Run 2 & Irradiated & $\beta$ & 0.018 $\pm$ 0.006 Hz/gram \\
Set A Run 2 & Control & $\beta$ & 0.008 $\pm$ 0.01 Hz/gram \\
\hline
\end{tabular}
\end{table}

Additionally, in the first Set A run, we warmed up the detectors to 50 K for approximately three days, before re-cooling the detectors and resuming taking data (see Fig. \ref{fig:warmup}). Both detectors saw a temporarily enhanced low energy event rate which decayed away with time. Shortly after cooling back down, the rate in the irradiated detector dropped below the trend line before the warm up, suggesting that this warm up period reconfigured the damage responsible for the enhanced low energy phonon event rate. The control detector appeared to asymptote towards the pre-warm-up trend line. 

In these datasets, we fit to a two-parameter power law with a time offset corresponding to the time at which the detectors were cold again. The power law parameter $\alpha$ appeared to be consistent between the irradiated and control detectors and with the fit $\alpha$ parameters before the warm-up period. In contrast to the pre-warm-up period, the post-warm-up prefactors $\beta$ appeared to be similar, with the irradiated detector's prefactor somewhat higher than the control detector. See table \ref{tab:fits_pwr_warmup} for the fit parameters.

\begin{table}[h]
\caption{\label{tab:fits_pwr_warmup} Time-offset power law (eqn. \ref{eqn:pwr_law_warmup}) fit parameters for the pre- and post-warm-up periods in Run 1 of the Set A detectors. Within uncertainties, the power law parameter $\alpha$ is consistent between the irradiated and non-irradiated detectors for both the pre- and post-warm-up periods, while the prefactor $\beta$ is reduced significantly for the irradiated detector while remaining relatively similar for the control detector.} 
\begin{tabular}{|c|c|c|c|}
\hline
Pre/Post Warm Up & Detector & Parameter & Fit Value\\
\hline
Pre Warm Up & Irradiated & $\alpha$ & 0.40 $\pm$ 0.06\\
Pre Warm Up & Control & $\alpha$ & 0.4 $\pm$ 0.2 \\
Pre Warm Up & Irradiated & $\beta$ & 0.038 $\pm$ 0.004 Hz/gram \\
Pre Warm Up & Control & $\beta$ & 0.005 $\pm$ 0.002 Hz/gram \\
\hline
Post Warm Up & Irradiated & $\alpha$ & 0.55 $\pm$ 0.07\\
Post Warm Up & Control & $\alpha$ & 0.44 $\pm$ 0.06 \\
Post Warm Up & Irradiated & $\beta$ & 0.0084 $\pm$ 0.0007 Hz/gram \\
Post Warm Up & Control & $\beta$ & 0.0062 $\pm$ 0.0005 Hz/gram \\
\hline
\end{tabular}
\end{table}

To directly compare the time-dependent rates seen in the Set A and Set B detectors, we assume that a fixed power law component $\alpha_{rad}$ describes the time dependence of the low energy phonon event rate in both the Set A and Set B detectors. We measured this parameter in the Set A irradiated detector in the pre-warm-up period in Run 1, where statistics are best. See table~\ref{tab:fits_pwr_fixed} for fit parameters.
\begin{eqnarray}
    R_f(t) = \beta t^{-\alpha_{rad}}
\end{eqnarray}

\begin{table}[h]
\caption{\label{tab:fits_pwr_fixed} Fixed exponent power law (eqn. \ref{eqn:pwr_law_fix}) fit parameters for the first and second run of the Set A detectors and for the set B detectors.} 
\begin{tabular}{|c|c|c|c|}
\hline
\textrm{Run} & Detector & Parameter & Fit Value\\
\hline
Set A Run 1 & Irradiated & $\beta$ & 0.038 $\pm$ 0.001 Hz/gram \\
Set A Run 2 & Irradiated & $\beta$ & 0.0121 $\pm$ 0.0008 Hz/gram \\
Set B & Irradiated & $\beta$ & 0.016 $\pm$ 0.006 Hz/gram \\
\hline
Set A Run 1 & Control & $\beta$ & 0.0047 $\pm$ 0.0004 Hz/gram \\
Set A Run 2 & Control & $\beta$ & 0.0031 $\pm$ 0.0007 Hz/gram \\
Set B & Control & $\beta$ & 0.0056 $\pm$ 0.0006 Hz/gram \\
\hline
\end{tabular}
\end{table}

\section{Detector Resolutions}
\label{appendix:resolutions}
As we previously described\cite{Run57Paper}, the energy resolutions of our detectors are limited by phonon shot noise which decreases in magnitude as a function of time, improving our energy resolution over time. While detailed studies of detector noise or resolution are beyond the scope of this paper, we provide benchmark phonon energy resolutions measured at specific times using a two channel, one amplitude phonon template optimal filter fitting 3.061 eV photon calibration data to broadly characterize our detector performance. 

\begin{table}[h]
\caption{\label{tab:resolutions} Resolutions and threshold measured in each detector on specific days of each run. } 
\begin{tabular}{|c|c|c|c|c|c|}
\hline
\textrm{Run} & Detector & Day & Phonon Resolution & Threshold ($\sigma$) & Threshold (eV)\\
\hline
Set A Run 1 & Irradiated & 12.2 & 312 $\pm$ 4 meV & 12 & 3.75 $\pm$ 0.05 eV \\
Set A Run 1 & Control & 12.2 & 285 $\pm$ 3 meV & 12 & 3.42 $\pm$ 0.03 \\
\hline
Set A Run 2 & Irradiated & 8.9 & 300 $\pm$ 9 meV & 12 & 3.6 $\pm$ 0.1 eV \\
Set A Run 2 & Control & 8.9 & 253 $\pm$ 6 meV & 12 & 3.04 $\pm$ 0.08 \\
\hline
Set B & Irradiated & 8.9 & 486 $\pm$ 2 meV & 12 & 5.83 $\pm$ 0.02 eV \\
Set B & Control & 8.9 & 429 $\pm$ 1 meV & 12 & 5.15 $\pm$ 0.02 \\
\hline
\end{tabular}
\end{table}

\begin{figure}[h]
\includegraphics[width=1\columnwidth]{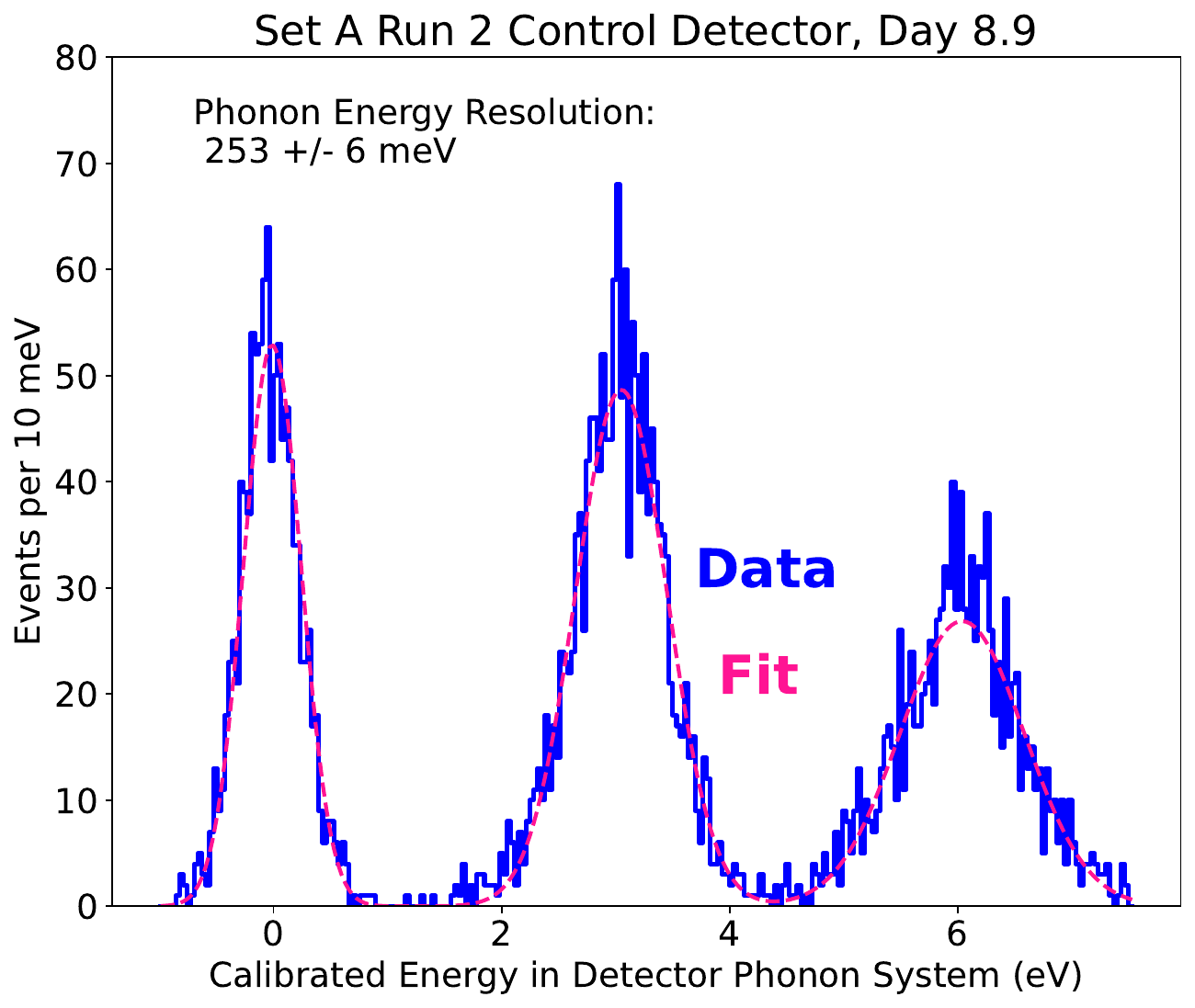}
\caption{\label{fig:calibration)} An example calibration spectrum from the second run of the Set A control detector, taken on day 8.9 of the run.}
\end{figure}


\end{document}